\documentclass{emulateapj}
\usepackage{graphicx}    
\usepackage{color}
\usepackage{natbib}

\newcommand{\Msun}{M_{\odot}}
\newcommand{\kms}{{\rm km} \, {\rm s}^{-1}}
\newcommand{\Mvir}{M_{\rm vir}}
\newcommand{\Rvir}{R_{\rm vir}}

\newcommand{\NHI}{N_{\rm HI}}
\newcommand{\Msh}{M_{\rm sh}}
\newcommand{\CF}{f_{c}}
\newcommand{\CFR}{f_{c}(<\hspace{-0.7 ex}R)}
\newcommand{\MGII}{Mg$\,$II }

\submitted{}

\begin{document}
\title{Orbiting Circum-galactic Gas as a Signature of Cosmological Accretion}
\author{Kyle R. Stewart\altaffilmark{1},       Tobias Kaufmann\altaffilmark{2},  James S. Bullock\altaffilmark{3,4},
        Elizabeth J. Barton\altaffilmark{3,4}, Ariyeh H. Maller\altaffilmark{5}, J\"urg Diemand\altaffilmark{6},
        James Wadsley\altaffilmark{7}}

\altaffiltext{1}{NASA Postdoctoral Program Fellow, Jet Propulsion Laboratory, Pasadena, CA 91109, USA}
\altaffiltext{2}{Institute for Astronomy, ETH Zurich, CH-8093 Zurich, Switzerland}
\altaffiltext{3}{Center for Cosmology, Department of Physics and Astronomy, The University of California at Irvine, Irvine, CA, 92697, USA}
\altaffiltext{4}{Center for Galaxy Evolution, Department of Physics and Astronomy, The University of California at Irvine, Irvine, CA, 92697, USA}
\altaffiltext{5}{Department of Physics, New York City College of Technology, 300 Jay St., Brooklyn, NY 11201, USA}
\altaffiltext{6}{Institute for Theoretical Physics, University of Zurich, 8057, Zurich, Switzerland}
\altaffiltext{7}{Department of Physics and Astronomy, McMaster University, Main Street West, Hamilton L85 4M1, Canada}

\begin{abstract} {
We use cosmological SPH simulations to study the kinematic signatures of cool gas accretion onto a pair of well-resolved galaxy halos.
Cold-flow streams and gas-rich mergers produce a circum-galactic component of cool gas that generally orbits with high angular momentum about the galaxy halo before falling in to build the disk.
This signature of cosmological accretion should be observable using background-object absorption line studies as features that are offset from the galaxy's systemic velocity by $\sim 100$ km/s.
Accreted gas typically co-rotates with the central disk in the form of a warped, extended cold flow disk, such that the observed velocity offset is in the same direction as galaxy rotation, appearing in sight lines that avoid the galactic poles. This prediction provides a means to observationally distinguish accreted gas from outflow gas: the accreted gas will show large one-sided velocity offsets in absorption line studies while radial/bi-conical outflows
will not (except possibly in special polar projections).
This rotation signature has already been seen in studies of intermediate redshift galaxy-absorber pairs; we suggest that these observations may be among the first to provide indirect observational evidence for cold accretion onto galactic halos.   Cold mode halo gas typically has $\sim3-5$ times more specific angular momentum than the dark matter.  The associated cold mode disk configurations are likely related to extended HI/XUV disks seen around galaxies in the local universe. The fraction of galaxies with extended cold flow disks and associated offset absorption-line gas should decrease around bright galaxies at low redshift, as cold mode accretion dies out.
}
\end{abstract}

\section{Introduction}
\label{Introduction}
Arguably the most substantive change in our theoretical picture of galaxy formation in the last decade has
involved the deposition of gas into galaxies.
Cosmological simulations set within the LCDM framework show that the majority of galactic baryons should be
accreted in a  ``cold mode'', which does not
shock heat in the halo prior to building the galaxy itself
\citep{Keres05, DekelBirnboim06, Brooks08, Dekel09, Keres09, FGKeres10, Stewart11a, vandeVoort11, FG11}.
This is quite different from the canonical picture where gas first obtains the halo virial
temperature and spin before falling in to form an angular momentum supported disk
\citep{Silk77, WhiteRees78, BarnesEf87, WhiteFrenk91, MallerBullock04}.  Instead,  short
cooling times prevent the development of a stable shock near the virial radius in halos that are
less massive than a critical threshold \citep{Binney77, BirnboimDekel03}
and cool gas gets deposited directly into galactic halos.

Previous simulation studies have noted that this gas seems to be rotating with large amounts of angular momentum,
which may help explain what happens to the significant potential energy of the infalling cool gas
\citep{Keres09,KeresHernquist09,Agertz09,Brook10}.  Here, we study the cold mode gas and its detectability in detail,
demonstrating that this accreted cool gas orbits about the halo before falling
in to build the central disk, delivering not just
fuel for star formation but also angular momentum to shape the outer galaxy.
This angular momentum enables an observable prediction:
cosmological accretion should be discernible in halo absorption-line studies as cool components
that are offset from the galaxy's systemic velocity
by $\sim50$ to $200 \, \kms$, co-rotating with the central disk in most cases.

Though the cold mode accretion prediction seems to be fairly robust and well understood analytically,
there remains no smoking-gun observational evidence that cold mode accretion is actually occurring in
nature. It is important to look for such a signature because it is always possible that additional
physics not currently included in simulations changes the story.  Such a situation is not out of the
question, given that the same simulations that are used to predict and characterize cold mode behavior
also over-predict the baryonic mass function of galaxies significantly \citep{Keres09b}. In order to
reconcile the relatively high level of baryonic deposition into galaxies with a correspondingly small
fraction of baryons locked up in galaxies today \citep{Guo10, Behroozi10}, one must appeal to outflows
of some kind.  Indeed, observations of cool halo gas around galaxies (typically via metal line absorption
systems) have emphasized the presence of gas outflows from galaxies
\citep[e.g.,][]{Steidel96, Shapley03, Martin05, Weiner09, Steidel10, Rubin10}.

 The clear observational evidence for outflow (rather than inflow) is at least understandable at high redshift;
 gas accretion at $z > 2$ is expected to flow along
 particularly dense filaments that can penetrate into the center of halos, even above
 the aforementioned critical halo mass. Dense, narrow streams of this kind
 may result in globally small covering fractions, making them difficult to observe
 \citep{FGKeres10, Kimm10,Fumagalli11}. Galaxies at $z \sim 2$ are also at the peak of
 cosmological star formation \citep{HopkinsBeacom06}; at these epochs, star-formation driven feedback processes
are even more likely to obscure any indicators of gas accretion (though feedback processes will invariably obscure
gas accretion signatures at all epochs, to degrees that are largely unknown).  Nevertheless, even at moderate redshifts
($z < 1.5$) there are no  definitive observational indications of significant cosmological gas accretion
 into galaxies in the form of cold mode accretion.

In a companion paper \citep{Stewart11a}  we showed that while the covering fraction of
cool accreted material remains relatively high in low-mass halos at moderate
redshifts ($\sim30-50\%$ within $R<50$ kpc at $z\sim1$),
the covering fraction of such gas should drop sharply in halos
more massive than the critical threshold for sustaining virial shocks
($\Mvir \sim 10^{12}\Msun$).  In principle, one would like to examine quasar-galaxy
(or galaxy-galaxy) pairs over a range of galaxy masses in order to witness
the end of the cold flow regime galaxy-by-galaxy.  But in order to do this, one must
differentiate absorption signals from accreted gas and absorption from gas
that has been deposited there from outflows.   In this paper we show that the kinematics
of absorbing gas can provide a means to distinguish its class: halo gas
from cold mode accretion generally rotates about the galaxy.

It should not be surprising that cold-mode gas tends to rotate with the disk.  This is the
primary mode by which disk galaxies are built in simulations
\citep[e.g.,][]{Dekel09}.  Cold flows and gas-rich mergers not only supply mass to galaxies
but also their angular momentum.  In fact, we find that cool accreted
gas tends to have higher and more coherent specific angular momentum than the dark matter in
the halo and this is one property that enables an observable kinematic
signature.

In the next section we describe our two simulations, each of which tracks the evolution of a
Milky Way-size galaxy down to $z = 0$, and we detail our method for creating
mock absorption sight lines through the simulations. We briefly explore the general properties
of the cool gaseous halos of our simulated galaxies in \S\ref{galaxyproperties}.
In \S\ref{kinematics}, we analyze one of the simulated galaxies at a single point in time,
detailing a unique observational signature associated with orbiting halo gas,
in the form of an extended disk of cool gas that co-rotates with the galaxy.
We investigate the statistical significance of this result \S\ref{vstime}, analyzing both simulations
over a broad range in redshift. We compare this co-rotation signature to observations
of galaxy-absorber pairs, and discuss the implications of our results in \S\ref{observations}.
We summarize our findings and conclude in \S\ref{conclusion}.

\begin{figure*}[tb]
  \includegraphics[width=0.99\textwidth]{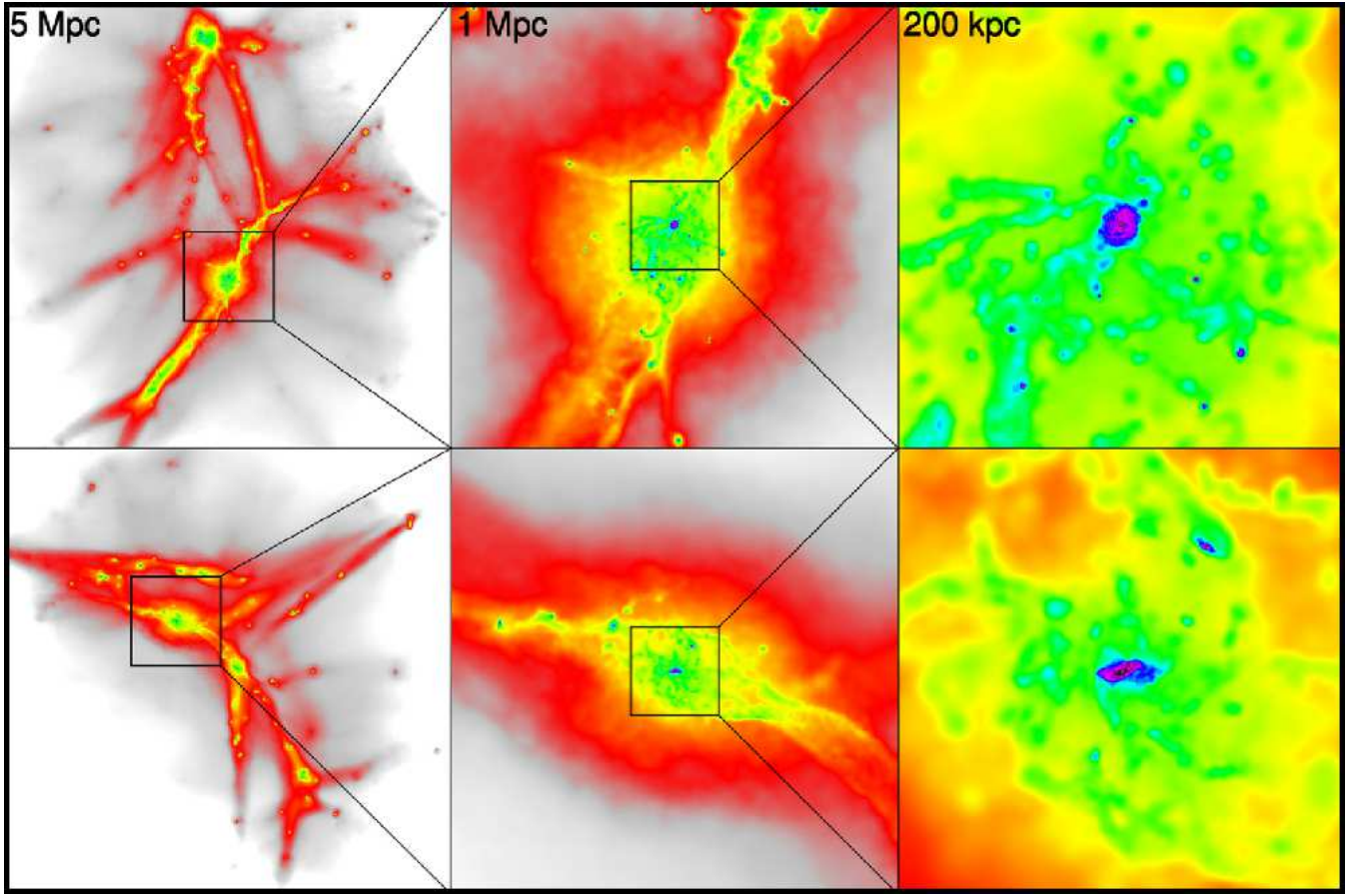}
  \caption{A visualization of projected gas density in each of our simulations at $z=2$, with the
  color variations on a log scale with red, green, and blue corresponding to gas column densities of
  $\sim0.03$, $0.4$, $2.0$ $\Msun/$pc$^2$, respectively.  The
  top panels show Halo $1$ while the bottom panels show
  Halo $2$.  From left to right, panel widths correspond to $5$ Mpc, $1$ Mpc, and $200$ kpc
  in co-moving coordinates.
  Note that both of our main galaxies reside within an extended filamentary structure at this redshift.
 }
  \label{zoommontage}
\end{figure*}

\section{Simulations and Analysis}
\label{simulation}
\subsection{Simulations}
Each of our two simulations utilize a separate set of  cosmological initial conditions, tracking the evolution of
a Milky Way size dark matter halo until $z=0$.  Because we primarily investigate the single most massive galaxy in each
simulation, we refer to these simulations throughout this paper as ``Halo $1$'' and ``Halo $2$.''
Halo $1$ has an active merger history until $z\sim1.5$, but is subsequently quiescent
(a dark matter only simulation of the same initial conditions was performed at very high resolution in
the Via Lactea II simulation of \citet{VL2}\footnote{The merger history and other data about this halo can be
found at $\texttt{http://www.ucolick.org/\char`\~diemand/vl/}$ }).
In contrast, Halo $2$ experiences a relatively quiescent early history,
but has a major merger at $z\sim0.8$, and another large merger at $z\sim0.2$.
For both simulations, the most massive galaxy has a dark halo virial mass of $\Mvir=1.4\times 10^{12} \Msun$ at $z=0$.

We use ``zoom-in'' multi-mass particle grid initial conditions generated
with the GRAFIC-2 package \citep{Bertschinger01}. A periodic box of $40$
co-moving Mpc is used in each simulation to properly account for
large-scale tidal torques. Each simulation contains a high resolution
region, limited to a $\sim6$ co-moving Mpc cube; in this region, the
masses of the simulation particles in the initial conditions are
$(m_{\rm dark}, m_{\rm gas}) = (17,3.7)\times10^5
\Msun$.\footnote{ For comparison,
recent simulations by \cite{KeresHernquist09} and \cite{FGKeres10}
have particle masses of $9\times10^4 \Msun$ and $4\times10^4 \Msun$, respectively.}

 We adopt the best-fit
cosmological parameters of the WMAP three-year data release \citep{WMAP3}:
$\Omega_{M} = 0.238$, $\Omega_{\Lambda} = 0.762$, $H_{0}= 73 $km s$^{-1}$
Mpc$^{-1}$, $n_s=0.951$, and $\sigma_8=0.74$

We use the code GASOLINE \citep{GASOLINE}, which is a smoothed
particle hydrodynamics (SPH) extension of the pure $N$-Body gravity
code PKDGRAV developed by \citet{Stadel01}. We use a gravitational force
softening of $332$ pc, which evolves co-movingly until $z = 9$ and
remains fixed from $z = 9$ to the present. The SPH smoothing length
adapts to always enclose the $32$ nearest gas particles and has the
minimum set to $0.05$ times the force softening length. Renderings of
the two simulations are shown in the upper (Halo 1) and lower (Halo 2)
panels of Figure \ref{zoommontage}. The panels from left to right show progressively
zoomed renderings of the gas density in each simulation at $z = 2$.

The code assumes a uniform UV background from QSO, implemented following
\citet{HaardtMadau96} and F. Haardt (2002, private communication).  It includes star
formation and Compton and radiative cooling
as described in \citet{Katz96}.  The code calculates the abundance of neutral hydrogen
by assuming  an optically thin ideal gas of primordial
composition and in ionization equilibrium  with the UV-background, treating collisional
ionization, photoionization and recombination processes.
Rates for collisional ionization from \citet{Abel97}, radiative recombination from \citet{Black81}
and \citet{Verner96} have been used.

In this paper, we focus on cool gas accretion and the possibility of detecting it as quasar absorption
systems, which primarily utilize metal lines (MgII, CIV, OVI, etc.).  To correctly produce
metal lines in a simulation of this kind requires
radiative transfer, metal diffusion and modeling of local ionizing sources, which are not included here.
Since we are only focusing on the qualitative behavior of halo gas, we will instead give results in terms of HI column
density, calculated in the optically thin limit, without local sources.
For column densities below the Lyman limit ($2 \times 10^{17}$ cm$^{-2}$) our computed column densities should be fairly robust.
For higher column densities, we expect a full treatment to lead to quantitative, but not qualitative differences.
(See \S\ref{cfvsr} for a quantitative comparison of our covering fractions with
the recent simulations by \cite{FGKeres10} and \cite{Fumagalli11} that do contain radiative
transfer treatments.)

The ``blastwave'' supernova feedback model implemented in our simulations
creates turbulent motions in nearby gas particles that keeps them from cooling and
forming stars, as described in \citet{Stinson06}.
The only free parameters in the star formation and feedback model
(minimum density threshold, $\rho_{\rm min}=0.1$ cm$^{-3}$; star formation efficiency factor,
$c^{\star}=0.05$; and fraction of supernova energy that couples with the ISM, $\epsilon SN=0.1$)
have been motivated by \citet{Governato07} in order to produce galaxies with a
realistic star formation rates, disk thicknesses, gas turbulence, and Schmidt law
over a range in dynamic masses.
The feedback model we use here is similar to those used in recent
simulations that have shown great success in producing realistic disk-type galaxies
\citep{Governato08}, matching the mass-metallicity relation \citep{Brooks07}, and matching the
abundance of Damped Ly$\alpha$ systems at $z=3$ \citep{Pontzen08}.  We
refer the reader to \citet{Governato07} for a more detailed description of the simulation code.

With the resolution implemented here, this feedback model results in minimal winds
($\sim100$ km/s) that mostly affect hot gas, and are more prominent in
halos with $\Mvir\lesssim10^{11}\Msun$ \citep{Shen10}.
Although outflowing gas may be more prominent
in higher resolution simulations with higher density thresholds and modified star formation and
supernova efficiency parameters \citep[e.g.,][]{Governato10},
we emphasize that the halos discussed here do \emph{not} have
outflows of cool gas that would be detectable through absorption.

We define the virial radius for our halos by \citet{BryanNorman98}, noting
that this is a fairly typical definition used in constructing merger histories from
$N$-body simulations of dark matter substructure \citep[e.g.][]{Stewart08}.
While this paper primarily focuses on mock observations of absorption lines through
our simulated galaxy halos, we analyze the formation and evolution of the galaxies themselves
in greater detail in an upcoming paper.
Both galaxies are disks before they reach the transition mass for hot mode accretion ($\Mvir\sim10^{12}\Msun$).
After this transition, both galaxies lose their reservoirs of cool high angular momentum halo gas.
Halo $1$ grows a massive bulge as hot gas cools from the halo, seemingly due to secular instabilities
(though it is possible that numerical resolution issues may give rise to artificial bulge-creation in the central regions, as well).
Halo $2$ develops into a more spheroidal system due to a major merger just prior to reaching the transition mass.
We note that previous works that have produced disk galaxies at $z=0$ using a similar code as the one used here
typically involve halos less massive than those we study here
\citep[e.g., $7\times10^{11}\Msun$ in][]{Governato08}, which may be the cause of these morphological differences,
since more massive galaxies are known to host a smaller fraction of disk galaxies \citep[e.g.][]{Weinmann09}.

\begin{figure}[tb!]
 \hspace{-2 em}
 \includegraphics[width=0.51\textwidth]{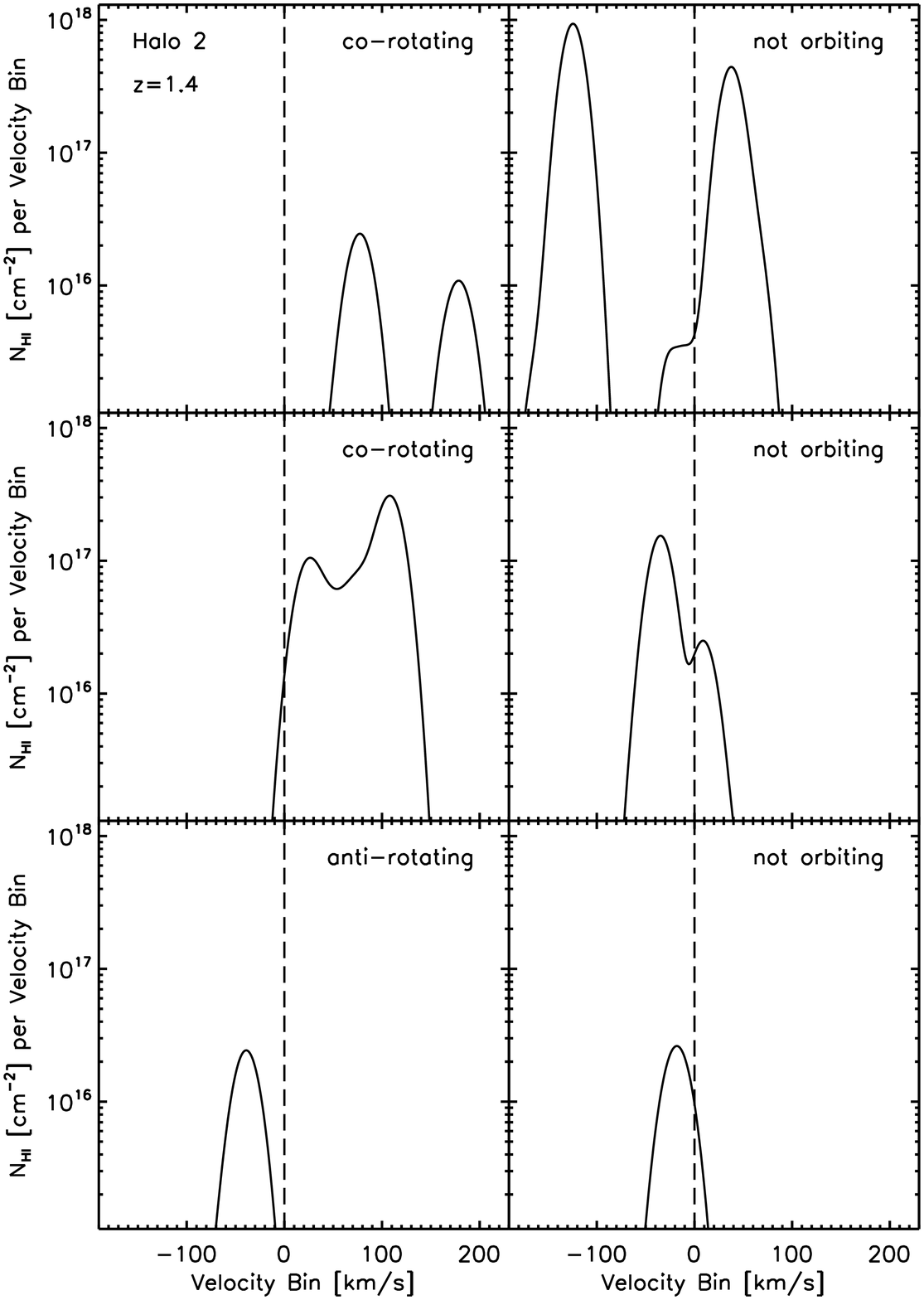}
 \caption{Example absorption sight lines taken from Halo $2$ at z=1.4, showing HI column density per unit
 velocity versus line of sight velocity.  In each panel, the galaxy systemic
 velocity is $0$ km/s, indicated by the vertical dashed line.
 \emph{Left:} sight lines that show orbiting gas, with velocity offsets from systemic
 (see \S \ref{absorptionlines} for definitions).  The top and middle plots are co-rotating, due to their spatial positions with respect
  to the galaxy, while the bottom panel is anti-rotating (i.e., the central galaxy's rotation curve would be positive on this side of the galaxy).
  The impact parameters of these particular sight lines are (from top to bottom): 50, 20, 60 co-moving kpc.
  \emph{Right:} sight lines that are not orbiting the galaxy, but show velocities near (or on both sides of) systemic.
  The impact parameters of these sight lines are (top to bottom): 6, 20, 40 co-moving kpc.
   }
\label{LOS_all}
\end{figure}

\subsection{Mock Absorption Sight Lines}
\label{absorptionlines}
In order to study the kinematics of cool halo gas in a manner useful for comparison to observations,
we construct regularly spaced lines of sight around our galaxies for a variety of viewing angles and for different times.
For each sight line, we use the analysis software TIPSY's ``absorb'' command to compute the column density
of neutral hydrogen per unit velocity, as a
function of velocity along the line of sight, mimicking the practice of searching for absorption gas along a quasar line of sight.
The column density is computed by first dividing each sight line into spatial bins of order the size of the spatial
resolution of the simulation.  For each spatial bin, each gas particle with a smoothing length that intersects the line of sight
contributes to the total mass density of the bin.  For each of these particles, the contribution of the spline kernel
to the column density of each bin that overlaps the kernel is integrated along the line of sight.  The velocity of each bin is
the mass-weighted velocity of particles contributing to the total column density of neutral hydrogen.
Each of these bins are broken into a number of sub-bins, with gas velocity and temperature linearly extrapolated from
adjacent bins.  Velocity profiles for each sub-bin is based on thermal broadening and dampening wings, with each sub-bin's
profile integrated over the velocity bin to determine its contribution.

We limit our examination to absorption lines with $\NHI>10^{16}$ cm$^{-2}$,
which allows us to focus on the fairly low column densities that
our simulations are most capable of reproducing\footnote{We again emphasize that comparing directly to
observed metal lines would require radiative transfer, metal diffusion and modeling of local ionizing
sources, which are not included here.  This is why we instead focus on the properties of HI gas at
column densities below the Lyman limit ($2 \times 10^{17}$ cm$^{-2}$).}.
This range also facilitates our later comparison with weak \MGII absorbing gas,
as it is approximately the minimum HI column density associated with \MGII detections \citep{Churchill00, Rigby02}.

Figure \ref{LOS_all} shows six example sight lines from our simulations (for Halo $2$ at $z=1.4$)
that meet our $\NHI>10^{16}$ cm$^{-2}$ criterion.
In all panels, we plot the column density of neutral hydrogen per velocity
bin as a function of velocity, with the galaxy systemic velocity set to $0$,
represented by the vertical dashed line.
In order to compare the kinematics of the absorbing gas to the central galaxy,
we classify each line of sight as either \emph{orbiting} or \emph{not orbiting}.

We define a sight line to be \emph{orbiting} if $>90\%$ of the absorbing gas lies
entirely to one side of the systemic velocity of the galaxy, where the galaxy's
systemic velocity is the HI-weighted projected velocity through the center of the galaxy.
Examples of sight lines classified as orbiting are shown
by the left three panels in Figure \ref{LOS_all}.  If this criterion is not met,
then we classify the kinematic signature as \emph{not orbiting}.
Note that this is a fairly conservative classification, in the sense that we
classify even ambiguous cases as not orbiting.
In the top right, the classification of \emph{not orbiting} is
due to the presence of two distinct components of gas that straddle the systemic velocity
(Interestingly, this configuration is reminiscent of something that might be tagged as
``outflow'' in an empirical study, though in our case, it is simply the chance
intersection of two clumps of accreted gas that happen to be orbiting in different
directions.) Such a configuration for is rare, however; more common
examples of gas that is not orbiting are the other right panels, each of which has
a single distribution of halo gas near the systemic velocity.
In contrast, the left middle panel shows some absorption on both sides of systemic,
but because $>90\%$ of the gas lies to a single side,
this sight line is still considered orbiting.  This configuration is also rare.
Single peaked, relatively narrow absorption
lines as shown in the two bottom panels are the most common profiles, with $>95\%$
of sight lines having a full width half maximum velocity spreads $<110$ km/s.\footnote{Because
our feedback model does not produce cool gas outflows, our absorption sight lines
are typically within $500$ km/s of systemic, and never offset by more than $1,000$ km/s.}

We note that in detail orbiting gas may not necessarily show a velocity offset from systemic along an
arbitrary projected line of sight.  Also, gas that does show a velocity offset from systemic
will have unknown radial velocities that could be substantially greater than the velocity offset detected.
Despite these shortcomings, we choose to characterize these velocity-offset sight lines as ``orbiting''
because a full $3$-dimensional analysis of our galaxies demonstrates that cool halo gas
detected by a velocity offset from systemic is typically orbiting the galaxy in a relatively
coherent fashion (see \S\ref{kinematics}).

If a sight line is found to be orbiting, we further classify it as either
\emph{co-rotating} or \emph{anti-rotating} by comparing the direction of the velocity offset
with the line-of-sight rotation of the galactic disk along the same projection
(sight lines to absorbing gas at near right angles to the galactic disk are
not included in this classification, as projection effects make these sight lines
ambiguous in terms of co-rotation or anti-rotation).
As an example, in the left panels of Figure \ref{LOS_all}, the top and middle panels are co-rotating,
due to their projected position with respect to the galaxy, while the bottom left panel is anti-rotating.

\section{Halo Gas properties}
\label{galaxyproperties}

Before using our simulated galaxies to probe kinematic signatures of cool halo gas, it is important to know the general properties of
our galaxies and their gaseous halos.  Because our primary focus is cool halo gas seen in absorption,
we choose to first investigate the radial extent of cool gas, to define a useful outer radius within which to focus
further analysis.

\begin{figure}[tb]
  \hspace{-2 em}
 \includegraphics[width=0.51\textwidth]{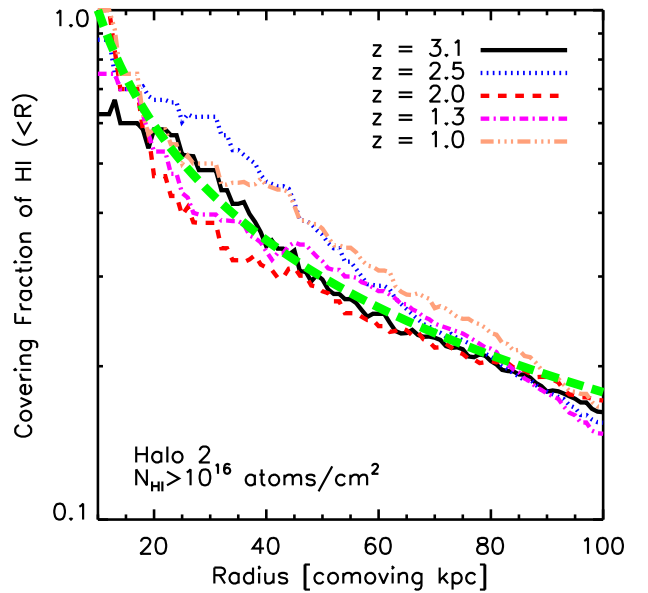}
  \caption{Example covering fraction versus projected radius profiles for Halo $2$ at various epochs.
  The radial profile of $\CFR$ is similar across a broad redshift range, so long
  as the galaxy is not experiencing a major merger.  The dashed green line shows a power law fit given by Equation \ref{eq:CFvsR1}.}
  \label{CFvsR}
\end{figure}

\subsection{Covering Fraction Versus Radius}
\label{cfvsr}

Some past studies of metal line absorption systems associated with galaxies suggest the presence of a critical radius for cool halo gas.
In this model, absorption gas extends in a roughly spherical configuration out to a given radius, and gas is unable to cool beyond this radius
\citep[e.g.,][]{Steidel95,TinkerChen08,ChenTinker08}.  Other studies suggest that the
covering fraction of detectable absorption systems should decreases as a power law in $R$ \citep{MartinBouche09,Steidel10}.

In order to quantify the radial extent of cool halo gas, we define the \emph{covering fraction} of accreted neutral hydrogen
as a function of radius, $\CFR$, as the total fraction of sight lines (within a projected radius $R$ from the center of the galaxy)
for which $\NHI>10^{16}$ cm$^{-2}$ (see \S \ref{absorptionlines} for details on absorption sight line construction).
Note that we include the galaxy disk in this analysis, so that $\CF\sim1$ for small values of $R$.  At large radii, this inclusion
of the galaxy to the total covering fraction is relatively minor ($<10\%,4\%,1\%$ of sight lines for $R<30, 50, 100$ co-moving kpc).

Figure \ref{CFvsR} shows five example profiles (for Halo $2$ at $z\sim3,2,2.5,1.3,1$)
of the covering fraction as a function of radius, from $R<10$ to $R<100$
co-moving kpc, averaged over three orthogonal orientations of the galaxy, for $\NHI > 10^{16}$ cm$^{-2}$.
Our simulated galaxies do not show a distinct truncation radius that cleanly separates a cold gas regime, but
instead we find a smooth progression from higher covering fractions
at small radii, to smaller covering fractions at larger radii (though this profile is less smooth during a gas-rich merger).
As shown in Figure \ref{CFvsR}, our $\CF$ vs. $R$ profiles are quite similar over a wide redshift range.  Each curve is well fit
by a power law in $R$, within the regime $10 < R < 100$, with $R$ in co-moving kpc:
\begin{equation}
\CFR=\left(\frac{R}{R_0}\right)^{-\beta}
\label{eq:CFvsR1}
\end{equation}
where $R_0$ represents an inner radius within which the covering fraction is $\sim100\%$.
The dashed green line in Figure \ref{CFvsR} is given by $R_0=10$ co-moving kpc and $\beta=0.7$.
We note that our results are in qualitative agreement with \cite{Steidel10}, who found
the covering fraction around LBG galaxies at $z\sim3$ to decline as a power law in $R$,
with a slope of $0.2-0.6$, very similar to our
value of $0.7$.  In addition, our results are also qualitatively similar to the lower redshift sample ($z<0.5$)
of \cite{Chen10b}, who found that the covering fraction is near unity at impact parameters $R\lesssim20$ kpc,
and declines roughly linearly in $\log R$ to $<5\%$ covering fractions at $R\sim100$ kpc.

In detail, the profile we present above also depends on the minimum value of
$\NHI$, with steeper radial profiles for higher column density gas \citep[similarly, higher equivalent with
absorption systems also show steeper radial profiles, see][]{Chen10b}.
Accounting for both simulated galaxies, we find more generally that the covering fraction as a function of
radius and minimum column density (from $\NHI\gtrsim10^{13}$ cm$^{-2}$ to $\NHI\gtrsim10^{18}$ cm$^{-2}$)
is well-fit to the following functional form\footnote{We
note that under the presumption of a power law form for $\CFR$, the \emph{differential} covering fraction of
gas \emph{precisely at} the projected radius R must follow the same power law slope, with a normalization that is lower by a
factor of $(1-\beta/2)$.}:
\begin{equation}
\CF(\NHI, <R)=\left(\frac{R}{R_0}\right)^{-\beta(\NHI)}  \, ,
\label{eq:CFvsR2}
\end{equation}
\begin{equation}
\beta(\NHI)=\beta_{16} + 0.1 \, \log[\NHI/10^{16} {\rm cm}^{-2} ]
\label{eq:beta}
\end{equation}
with $R_0\sim10$ co-moving kpc and $\beta_{16}\sim0.6-0.8$.
Because the cool gaseous halos of our galaxies
die out after cold mode accretion ends ($z<1.3$ and $z<0.8$ for Halo $1$ and Halo $2$, respectively) we caution that
this derived fit for $\CFR$ is based on a narrow halo mass and redshift range: $\Mvir\sim10^{11}-10^{12}\Msun$ and $3<z<1$.

We also caution that our simulations do not have treatment of radiative transfer (see \S\ref{absorptionlines}), which may
affect the detailed values of $\NHI$ presented here.  In order to give some estimate of how this may impact our resulting
covering fractions, we compare the covering fraction of our galaxies
at $z=2$ with recent high-z simulations of \cite{FGKeres10} and \cite{Fumagalli11}, both of which include post-processing
radiative transfer prescriptions for galaxies of similar mass to those analyzed here.  At $z\sim2$, these works both find that
$\CF(<2\Rvir)\sim4\%$ and $\CF(<\Rvir)\sim10\%$ for $\NHI>10^{17.3}$ cm$^{-2}$, while our simulations (utilizing the above best-fit
equation, and $\Rvir\sim250$ co-moving kpc at $z=2$) have covering fractions of $\sim4\%$ and $\sim7\%$, respectively,
for the same column densities.

This indicates that the lack of radiative transfer may lower the available HI
content in the halos of our galaxies, suppressing the covering fraction by a factor of $\sim30\%$ at a fixed column
density threshold. (This, in turn, suggests that our reported column densities may be slightly higher than they would
be with inclusion of radiative transfer).  Because the rest of this paper primarily discusses the kinematic properties
of cool halo gas in our galaxies, only utilizing the integrated column densities of sight lines to determine if they
are potentially observable or not, we do not believe the lack of radiative transfer will substantially alter our
findings.

It is worth noting that most of the accreted halo gas resides within $R<100$ co-moving kpc of our simulated galaxies,
with the only notable exception being major, gas-rich galaxy mergers that have not yet fallen within this radius.
\citep[Mergers of this kind are more likely at high redshift, when mergers are more common, and galaxies are more gas rich
than at lower redshift; see e.g.,][]{Stewart09a,Stewart09b}.
Since the infall time of major mergers is short \citep[e.g.,][]{BoylanKolchin08},
we will focus our analysis on $R<100$ co-moving kpc for the remainder of this paper.
For our simulated halos, this choice of radius corresponds to $\sim \Rvir/2$ at $z=3$ and $\sim \Rvir/3$ at $z=0$.

\subsection{Mass and Angular Momentum Evolution}

\begin{figure*}[f]
 \hspace{-2 em}
 \includegraphics[width=0.54\textwidth]{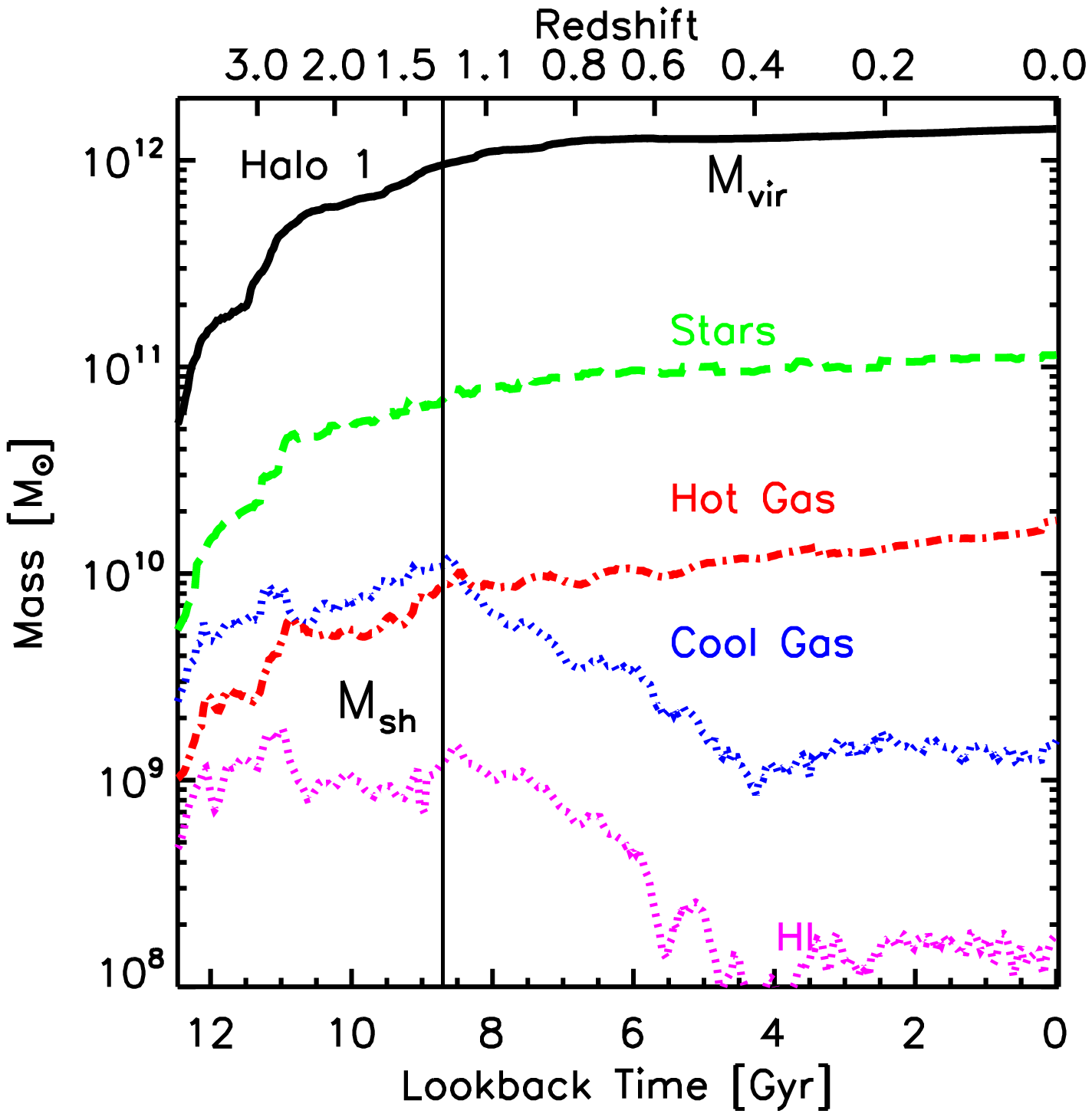}
 \hspace{-2 em}
 \includegraphics[width=0.54\textwidth]{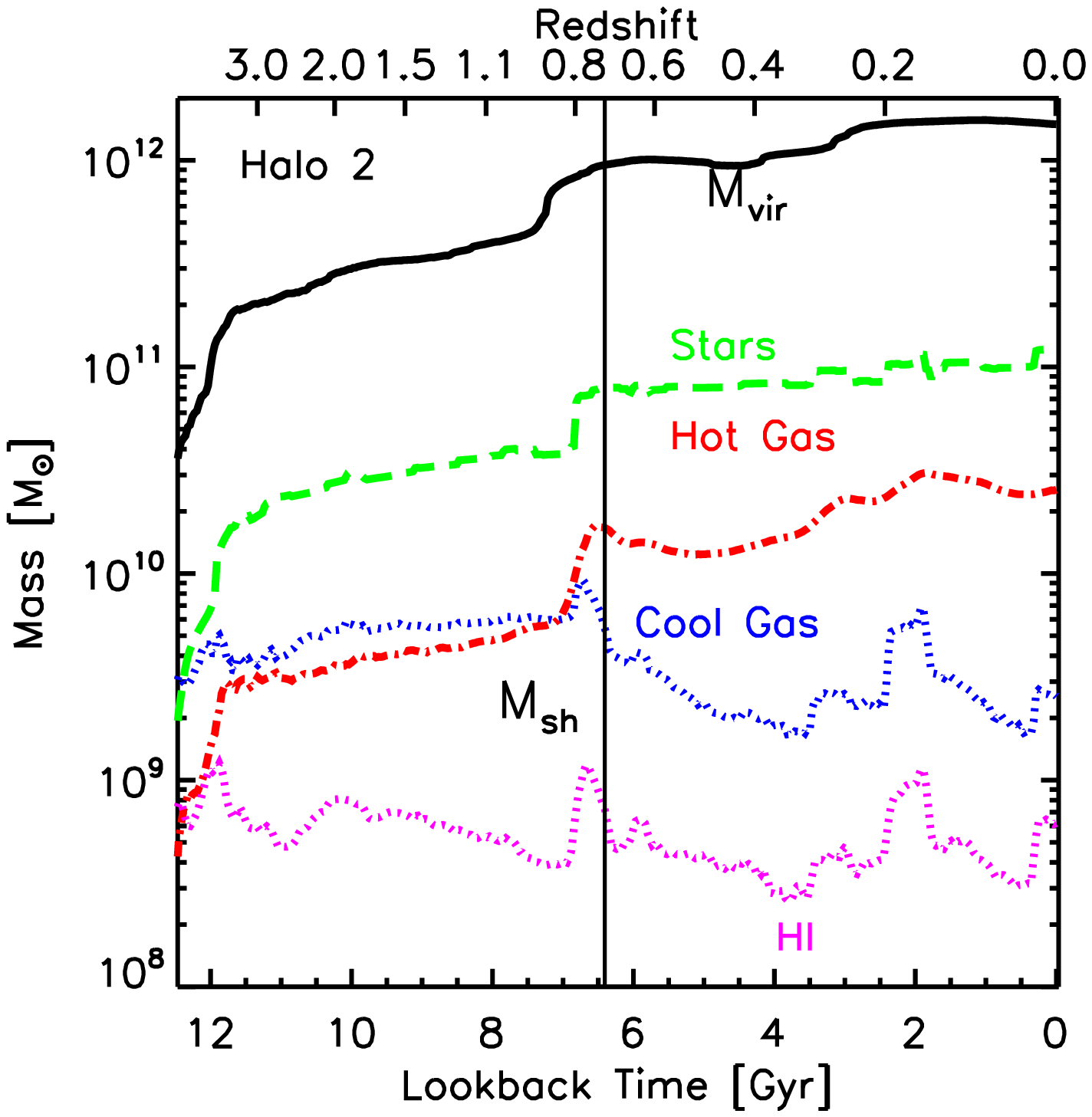}
 \\

 \vspace{-2 em}
 \hspace{-2 em}
 \includegraphics[width=0.54\textwidth]{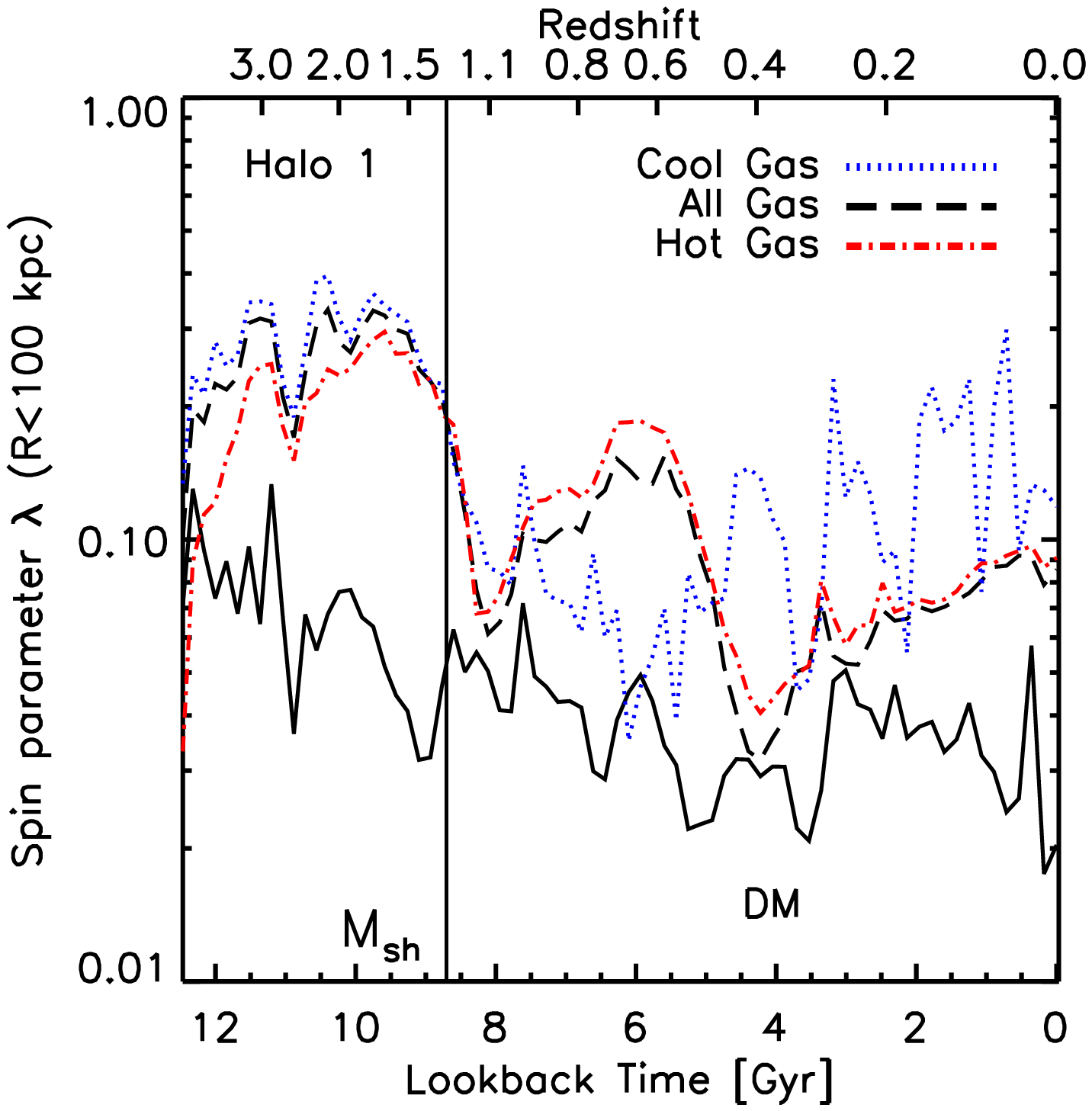}
 \hspace{-2 em}
 \includegraphics[width=0.54\textwidth]{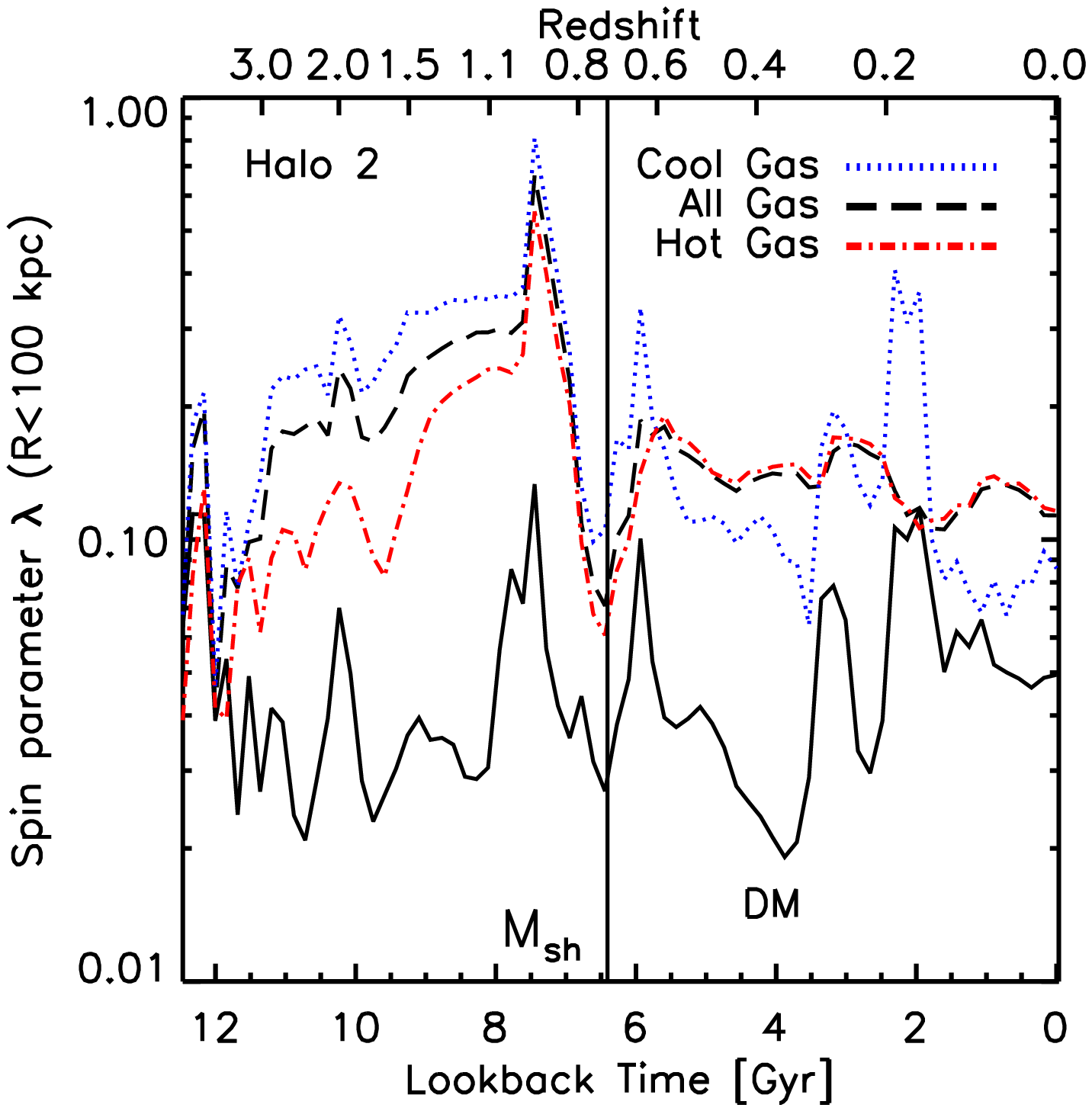}
 \caption{Baryonic masses and spin parameters ($R<100$ co-moving kpc of each
 simulated galaxy) versus time.  The left panels show Halo $1$, while the right panels show Halo $2$.
 \emph{Top:} the total mass within the virial radius is shown by the solid black
 line, while the stellar mass, hot gas mass ($T>10^5$ K), cool gas mass ($T\leq10^5$ K),
 and HI gas mass within $R<100$ co-moving kpc are given by the dashed green,
 dot-dashed red, dotted blue, and dotted magenta lines, respectively.
 \emph{Bottom:} the spin parameter of the dark matter and gas within $R<100$
 co-moving kpc (see Equation \ref{eq:spinparameter}).
 The dark matter spin parameter is given by a solid black line, while the spin
 parameters of cool gas, hot gas, and all gas are given by the
 blue dotted, red dot-dashed, and black dashed lines, respectively.  Note that after rapid accumulation of angular momentum from
 mergers, the dark matter spin parameter quickly settles to pre-merger values,
 while the gas spin parameters stay higher considerably longer.
 The vertical line in each panel shows where the galaxy crosses the transition
 from cold mode to hot mode accretion, $\Mvir=\Msh\sim10^{12}\Msun$,
 as motivated by \citet{DekelBirnboim06} for shocks at a moderately large fraction of the virial radius.  After this threshold,
 little cool gas reaches the galaxy from cold flow accretion.
 }
\label{masses}
\end{figure*}

The top panels in Figure \ref{masses}
show the mass buildup of each galaxy over cosmic time, with
the left and right panels showing Halo $1$ and Halo $2$, respectively.  In each panel, the solid black line shows the total virial mass
of each halo (including dark matter and all baryons within $R<\Rvir$), while the other curves
represent the baryonic masses within $R<100$ co-moving kpc; the dashed green, dot-dashed
red, dotted blue and dotted magenta lines show the stellar mass, hot gas mass ($T>10^5$ K), cool gas mass ($T\leq10^5$ K), and
HI gas mass, respectively.

Once each galaxy crosses the critical halo mass for forming shocks, $\Msh$,
accreted gas is shock heated strongly enough that
very little cool gas reaches the inner halos of our galaxies.   This point is
indicated by vertical lines in each panel of Figure \ref{masses}.
We see that after this critical transition, the gas mass in the halo becomes dominated by the hot component (red dot-dashed line)
and the cool component drops in comparison (blue dotted line).  At the same time, the
covering fraction of cool gas drops severely \citep[][]{Stewart11a} and the
galaxies begin to evolve into more bulge-dominated morphologies, rather than disks (see \S\ref{observations2} for more on this transition).

Given that the velocity distribution of halo gas is the key observational signature we are exploring, it is useful to characterize the
kinematics of orbiting halo gas in terms of its angular momentum.  A useful way to do this
is via a dissipationless spin parameter \citep{Peebles69}. Specifically, we
compare the specific angular momentum $j$ of the dark matter, hot gas, and cool gas
as a function of time in each halo using the spin parameter defined by \citet{Bullock01}
\begin{equation}
\lambda_x \equiv \frac{j_x}{\sqrt{2} \, V \, R} \,  .
\label{eq:spinparameter}
\end{equation}
Here $j_x$ is the specific angular momentum of component $x$ ($=$ dark matter, hot gas, cool gas, and all gas) within a sphere of radius $R$  and $V$ is the circular velocity measured at radius $R$.

The bottom panels of Figure \ref{masses} show the spin parameter for the dark
matter within $R=100$ co-moving kpc as a solid black line.  The
spin parameter evolution of the gaseous components within the same radius is represented by the black dashed (all gas), blue dotted (cool gas), and red dot-dashed lines (hot gas).   Note that the spin parameter in dark matter evolves somewhat stochastically, spinning up in response to mergers \citep{Vitvitska02, Maller02} but then settling down as the halos re-virialize and the angular momentum is redistributed \citep{Donghia07}. The redistribution does not happen as readily in gas,  which tends to cohere and spin up more strongly.  The average spin for the dark matter bounces around $\lambda_{\rm dm} \sim 0.04$ over time, at a value that is quite similar to the expectations of many past studies from dissipationless simulations \citep[recently,][]{Bett07, Maccio07, Berta08, Bett10}.
On the other hand, the gaseous halo components have much more angular momentum, demonstrating a more sustained response to mergers and inflow.  The cool gas has the highest
spin of all, some $\sim 3-5$ times higher than that of the dark matter with $\lambda_{\rm cool gas} \sim 0.1-0.2$.  We emphasize that this result is particularly important for theoretical studies that try to understand angular-momentum supported disk sizes.  There is long history of studies relying on spin parameter distributions derived from dissipationless simulations to predict disk sizes and bulge properties in galaxies \citep[e.g.,][]{FallEfst80, Blumenthal86,MoMaoWhite98, Bullock01}.  Given that cool halo gas is the primary component that actually builds galaxies \citep[e.g.,][]{Dekel09}, our result that $\lambda_{\rm cool gas} \sim 4 \, \lambda_{\rm dm}$ motivates  reconsideration of the issue of disk sizes in a cosmological context \citep[see also][]{MallerDekel02}.

\begin{figure}[tb!]
 \includegraphics[width=0.47\textwidth]{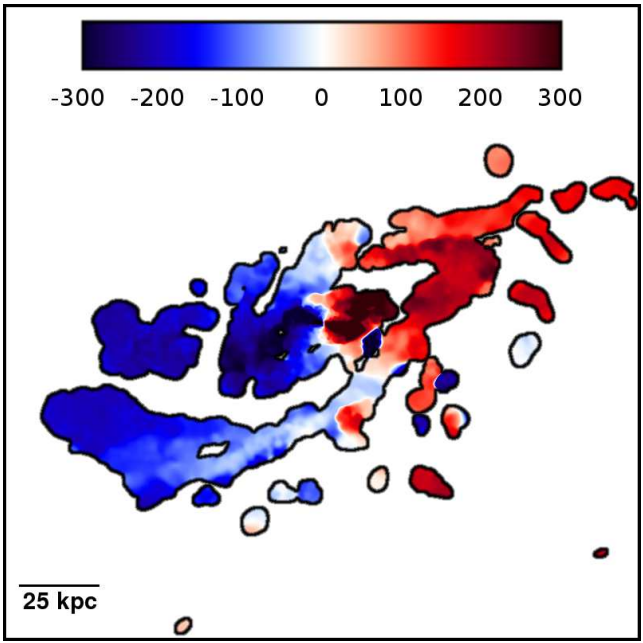}
 \caption{Line of sight velocity of cool halo gas in Halo $2$'s main galaxy at $z=1.4$.  The image width is $200$ co-moving kpc, and the
   galaxy is viewed near edge-on at an inclination of $75$.
   The black contours enclose regions with $\NHI>10^{16}$ cm$^{-2}$, and the blue-red shading shows the average (mass-weighted) velocity of HI along the line of sight,
   from $-300$ to $300$ km/s.
   The galaxy (inner region) tends to rotate in the same direction as cool halo gas, even out to $100$ co-moving kpc).
   }
\label{corot_pretty}
\end{figure}
\begin{figure*}[tbh!]
 \includegraphics[width=0.49\textwidth]{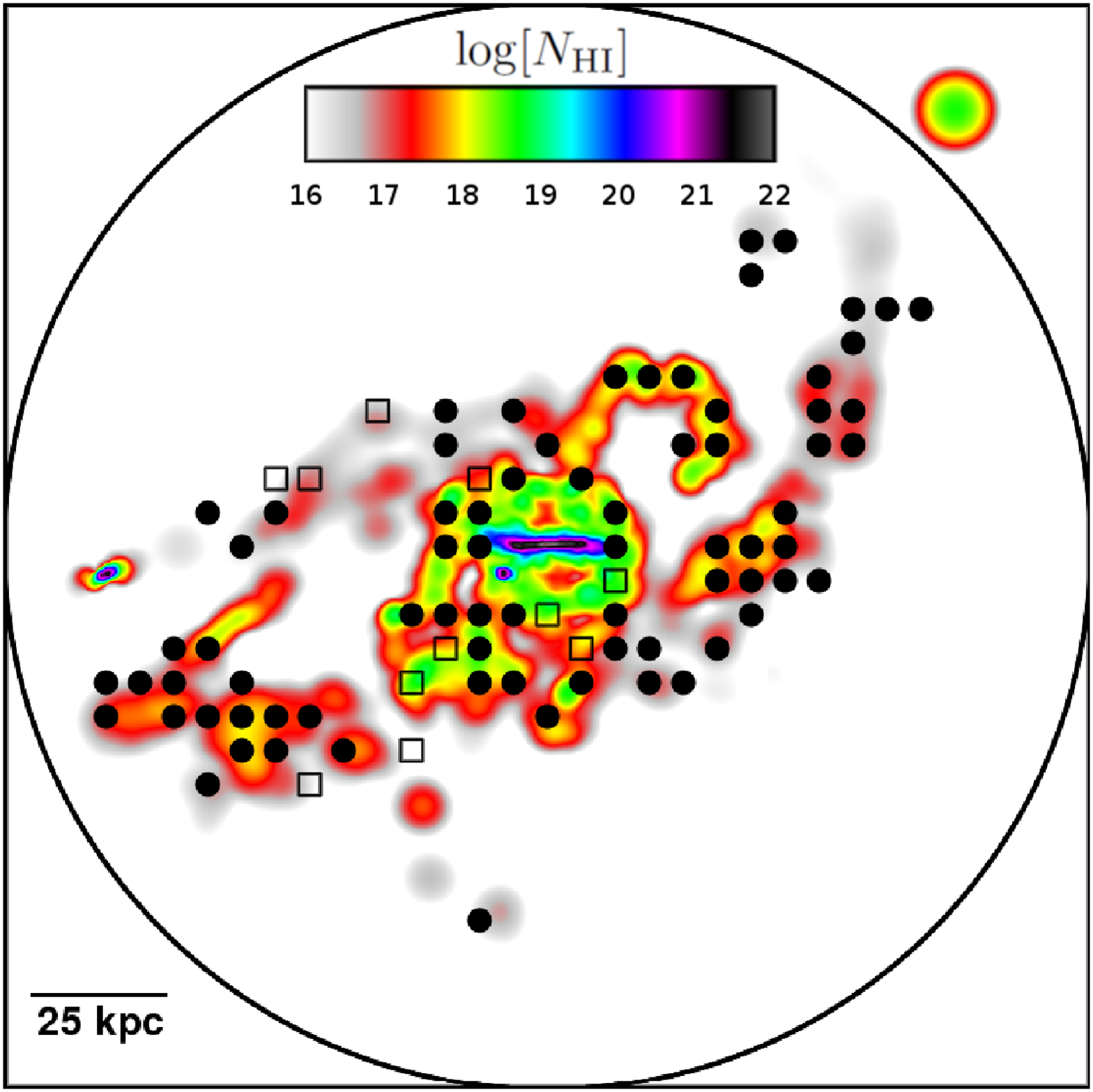}
 \includegraphics[width=0.49\textwidth]{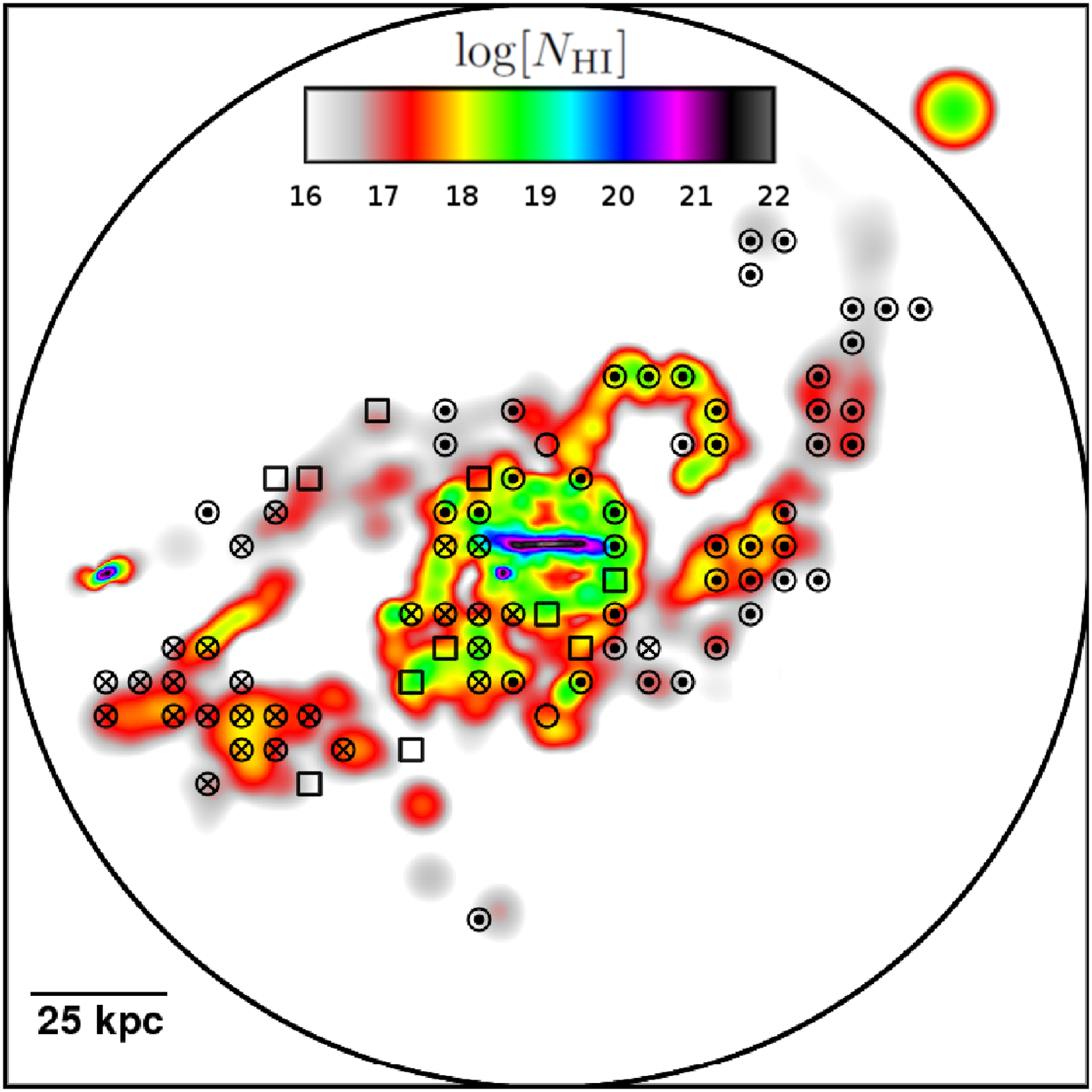}
 \caption{Co-rotation of cool halo gas in Halo $2$'s main galaxy at $z=1.4$.
   The large black circles denote a radius of $100$ co-moving kpc and the logarithmic color scale shows the projected column
   density of HI.  In the left panel, observable sight
   lines ($\NHI>10^{16}$ cm$^{-2}$) are categorized as orbiting (filled circles)
   or not orbiting (open squares).  In the right panel, the direction of
   the velocity offset of orbiting sight lines is explicitly displayed, with
   redshifts and blueshifts given by circle-X and circle-dot symbols, respectively.
   }
\label{corot_boxes}
\end{figure*}

Before moving on, we note that while we have included disk gas and halo gas in our $\lambda$ determinations,
the gas associated with the central disk typically constitutes only $\sim10-20\%$ of gas within this radius, and is therefore unimportant to the quantities plotted here.  That is, the spin parameters we are discussing are dominated by {\em halo gas}. We will present a more detailed study of the angular momentum buildup of these galaxies in an upcoming paper.

The relatively large amount of angular momentum in the cool component suggests that cold-accretion material may display a smoking-gun signature in its kinematics.  In the next section we demonstrate not only that this gas is rapidly orbiting about the halo center, but that it tends to rotate {\em with} the central disk.  This is not surprising given that this is the gas that fuels disk formation in the end.  What is important for our purposes is that  this co-rotation between cool
halo gas and the galaxy disk can be observed via halo absorption studies.

\section{Kinematics of Cool Halo Gas}
\label{kinematics}
\subsection{Co-rotation of Cool Accreted Halo Gas}
\label{kinematics1}
The kinematics of cool gas in the halos of galaxies serve as an important probe of galaxy formation.
Unfortunately, the detailed nature of cool halo gas is difficult to determine empirically, as
observations are presently limited to studying absorption along quasar lines of sight.
As a result, even when the resulting sight line is close enough to a foreground galaxy to
probe its gaseous halo, these observations only allow for a single absorption
line of sight per galaxy.  Fundamental properties such as the covering fraction and kinematics
of cool halo gas can only be accumulated in a statistical manner, for many different galaxies,
with varied properties, redshifts, and distances between the absorbing gas and the galaxy.
Despite these inherent difficulties, observations of cool halo gas have progressed substantially in the past decade---typically
utilizing \MGII absorption, which arises in low ionization, metal enriched gas of fairly high density
and probes neutral hydrogen column densities of $\sim10^{16-22}$ cm$^{-2}$ \citep{Rigby02, Churchill00}.

Still, it remains unclear what fundamental processes give rise to the absorbing gas.
Different observations suggest that it may
result from extended thick disks \citep{Steidel02}, cool cloud formation from hot halo gas \citep{ChenTinker08}, cool gas
that is outflowing from galaxy-scale winds \citep{Bond01, Bouche06, Prochter06, Weiner09,Steidel10,Rubin11},
tidal streams from recent merger activity \citep{Churchill05, Rubin10}, or a complex combination of the above,
with inclusion of cosmological gas accretion along filaments \citep{Kacprzak07, Kacprzak10}.
In this section, we create mock observations of our simulated galaxies, utilizing line of sight absorption profiles of cool gas
(as detailed in \S \ref{absorptionlines}) to probe observable kinematic signatures of cool accreted gas.

Before analyzing our mock absorption lines, we give a visual impression of
the rotation of cool gas within $R<100$ co-moving kpc in Figure \ref{corot_pretty}.  For each pixel in the image, the
blue--red shading corresponds to the mass-weighted average velocity of HI gas along the line of sight.
The galaxy, seen near edge-on in the inner region of the figure (at an inclination of $75$ degrees),
shows a clear rotation signature; gas on the
left side of the disk shows a negative (blue) line of sight velocity while the right side of the disk
has a positive (red) line of sight velocity.  As shown, the rotation signature extends well beyond the galactic disk; in fact, the
\emph{majority} of cool halo gas out to $R<100$ co-moving kpc is co-rotating with the galaxy.

This kinematic correlation
is not surprising, given that the angular momentum vectors of the inner and outer
regions of dark matter halos are expected to be in rough alignment
\citep[typical misalignment $\sim30$ degrees; e.g.,][]{vandenBosch02,SharmaSteinmetz05,Bailin05,Bett10}.
Nor is it surprising that cold mode gas has high angular momentum before falling in to build the disk,
as multiple past investigations into the properties of cold mode accretion have noted the presence of
high angular momentum cool gas in proximity to (but more extended than) the galaxy \citep{Keres09,KeresHernquist09,Agertz09,Brook10}
Indeed, the gravitational potential energy of infalling gas should result in velocities of $\sim100-300$ $\kms$,
and if this energy is not dissipated by shock-heating, the conversion of potential energy into rotational velocity
is a quite natural possibility \citep[see discussion in][]{Keres05}.

\begin{figure*}[tbh!]
 \hspace{-2.5 em}
 \includegraphics[width=0.37\textwidth]{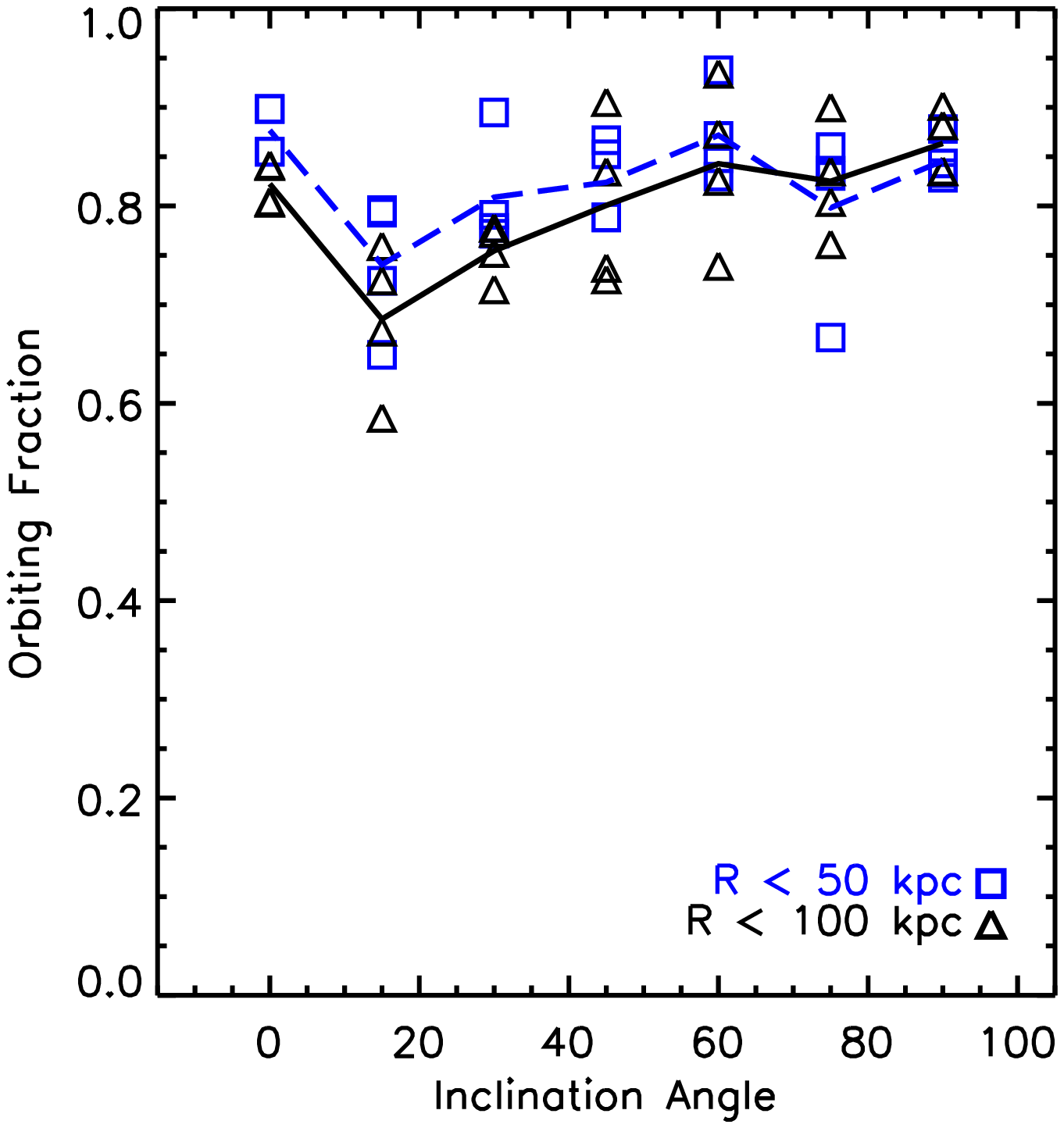}
 \hspace{-2.5 em}
 \includegraphics[width=0.37\textwidth]{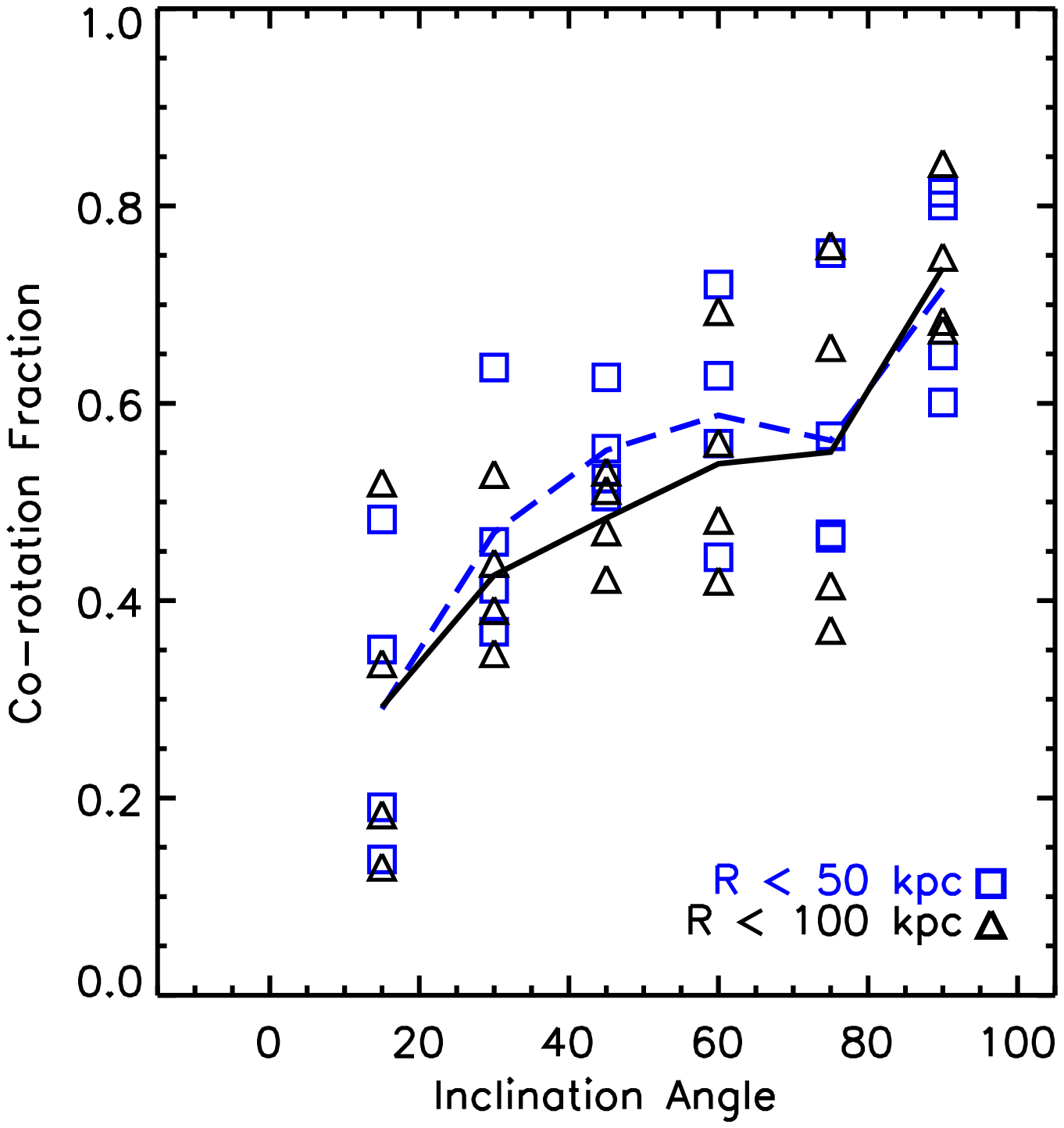}
 \hspace{-2.5 em}
 \includegraphics[width=0.37\textwidth]{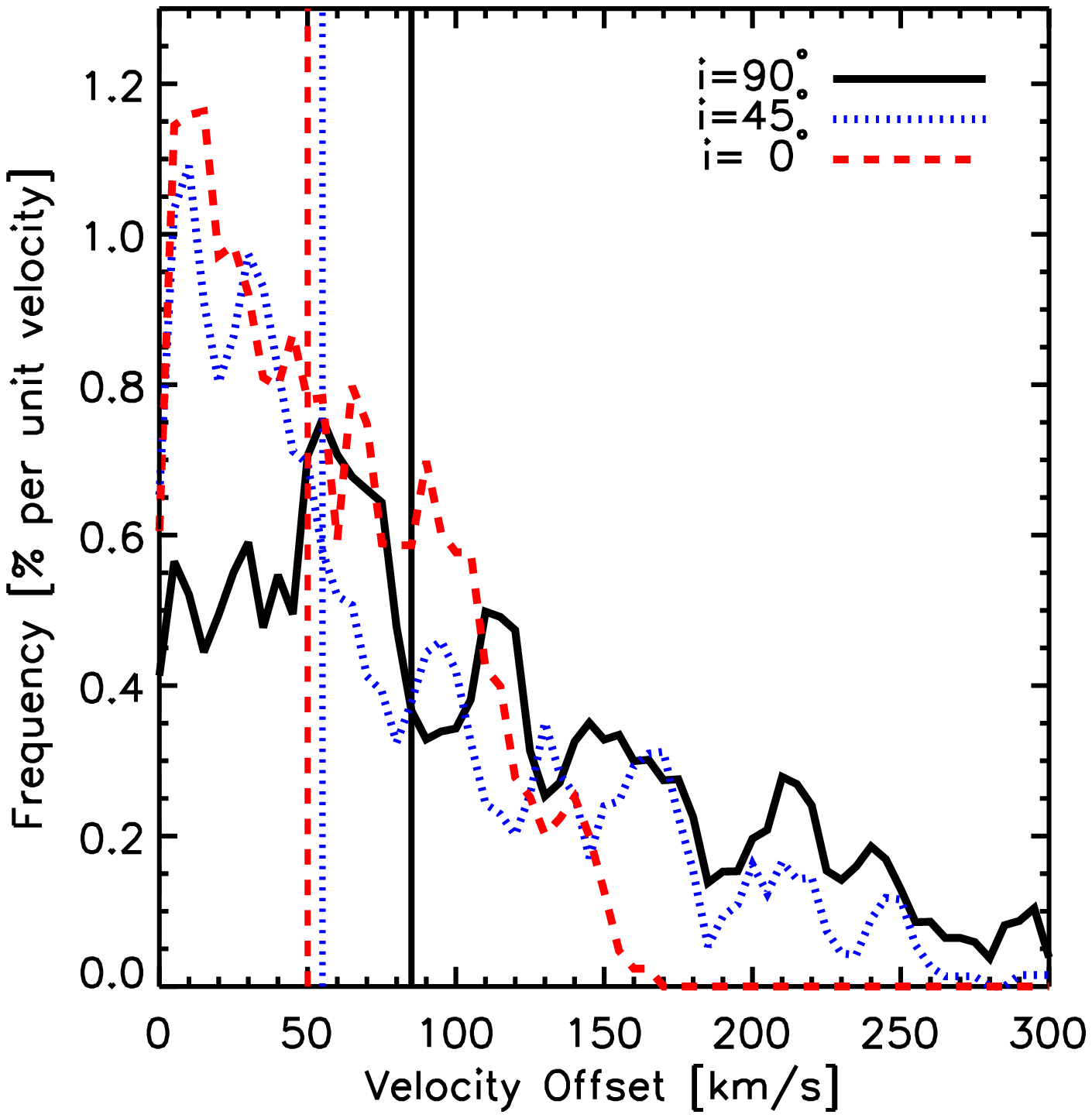}
 \caption{Observable characteristics of halo absorption line kinematics as a function of central galaxy inclination angle for Halo $2$ at $z=1.4$.
 At each angle, we have analyzed four viewing orientations around the galaxy.
 \emph{Left and Middle:} the orbiting and co-rotation fractions for different galaxy orientations.
 Squares and triangles show $R<50, 100$ co-moving kpc statistics,
 with each symbol representing a different orientation of the galaxy.
 The orbiting fraction is universally high, while co-rotation fractions vary strongly with galaxy inclination.
 \emph{Right:} magnitude of velocity offset with galaxy inclination.
 Each curve shows the relative frequency of different velocity offsets for different galaxy inclinations,
 for lines of sight with $R<100$ co-moving kpc and $\NHI>10^{16}$ cm$^{-2}$.  The vertical lines indicate the median velocity offset for each
 inclination angle.
 }
\label{corotvsangle}
\end{figure*}

While Figure \ref{corot_pretty} gives an intuitive visual impression of this rotation, a mass weighted velocity of cool gas
along a line of sight is not directly observable.  In order to determine if this co-rotation signature might be
observed via studies of absorption systems, we analyze $794$ regularly spaced absorption profile sight lines
within $R<100$ co-moving kpc of the central galaxy\footnote{The $794$ sight lines form a square grid of $32^2$ cells within a cube of length $200$ co-moving kpc per side, after removing cells with $R\geq100$ co-moving kpc.}.
Comparing each absorption profile to the rotation signature of the galactic disk, we categorize sight lines
as orbiting or not orbiting based on their velocity offsets, and further divide orbiting sight lines into those that are co-rotating or anti-rotating with the galactic disk
(see \S\ref{absorptionlines} for detailed definitions of these classifications).

Figure \ref{corot_boxes} again shows Halo $2$ at $z=1.4$ (this time viewed edge-on), but presenting a
more observationally-oriented method than before.
For both panels in Figure \ref{corot_boxes}, the image width is $200$ co-moving kpc,
with the black circle marking a $100$ co-moving kpc radius from the
center of the galaxy.  The colored contours in the background of each panel show the projected column density of HI
with the color code on a log scale from $>10^{16}$ cm$^{-2}$ (red) to $>10^{22}$ cm$^{-2}$ (purple/black).
On this scale, the edge-on disk is visible as the dark horizontal bar across the very center of the image.
We remove sight lines in the central region ($R<10$ kpc) from our analysis,
since we are interested in the properties of the cool \emph{halo} gas,
not gas that is likely associated with the galactic disk.  (This is why there are no symbols in the center of either image.)
Along this projection, the central galaxy is redshifted on the left and
blueshifted on the right.

In the left panel, we have overlaid symbols representing the division
between sight lines with significant velocity offsets (orbiting sight lines; filled circles)
and those that do not (open squares).
As the figure makes clear, the vast majority of sight lines are orbiting
about the central galaxy, with velocities offset from the systemic velocity of the system.
At this particular snapshot, the \emph{orbiting fraction} (fraction of all
$\NHI>10^{16}$ cm$^{-2}$ sight lines which are orbiting) is $\sim90\%$.

The right panel is identical to the left, except that orbiting sight lines
(filled circles) now have their velocity offset shown explicitly, so
that we may determine which are co-rotating or anti-rotating with the galaxy.
Sight lines that are redshifted with respect to systemic (``into'' the image) are shown by circle-X symbols,
while blueshifted sight lines (``out of'' the image) are given by circle-dots.
For this example snapshot and orientation, co-rotating sight lines follow
the kinematics of the galaxy: redshifted on the left side of the image or blueshifted
on the right side of the image.  The \emph{co-rotation fraction} (fraction of $\NHI>10^{16}$ cm$^{-2}$ sight lines
that show this co-rotation signature) is $\sim70\%$ for this image\footnote{Since lines
of sight perpendicular to the plane of the galaxy cannot be
classified as co-rotating or anti-rotating, we leave these as open circles,
and do not include them in computation of the co-rotation fraction.}.

\subsection{Effects of Galaxy Inclination}
\label{vsinclination}
In order to investigate the generality of our results, we repeat our kinematic analysis from \S\ref{kinematics1} for four equally spaced
viewing angles around each galaxy (at the same inclination, from $0-135$ degrees, inclusive) as well as seven equally spaced
galaxy inclinations ($0-90$ degrees, inclusive) for a total of $28$ unique orientations per galaxy.
Continuing with the $z=1.4$ snapshot of Halo $2$, Figure \ref{corotvsangle} shows how the orbiting fraction (left panel)
co-rotation fraction (middle panel) and velocity offset distributions (right panel) vary with these different orientations.
In the left two panels, the x-axis shows the $7$ different inclinations of the galaxy, while each of the $4$ symbols along the y-axis shows
one of the $4$ different viewing angles around the galaxy\footnote{The order of
rotation for all $28$ orientations is first by angle, then by inclination.}.
In order to explore how the orbiting and co-rotation fractions might vary with radius,
we show statistics for two choices of outer radii (squares, triangles for $R<50, 100$ co-moving kpc, respectively).
The dashed and solid curves gives the mean values at each inclination for $R<50$ and $R<100$ co-moving kpc.

Regardless of the galaxy's orientation, the overall orbiting signature stays constant, with
a mean orbiting fraction of $\sim70-90\%$.  This velocity offset
signature is \emph{not} a strong function of radius from the central galaxy.
Cool halo gas at $R=100$ co-moving kpc seems equally likely to be orbiting about the galaxy in some fashion as
gas that is closer in to the galaxy.

In contrast, the co-rotation fraction varies strongly with the orientation of the galaxy.
At a fixed inclination, the co-rotation signature shows scatter of $\sim20-30\%$, depending on the viewing angle, but
even more drastic is the dependence on inclination.  Averaging over the $4$ viewing angles, $\sim70\%$ of absorption sight lines
appear to be co-rotating with the galaxy when viewed near edge-on.  When viewed near-face on, not only does this co-rotation
signature drop, but the cool halo gas actually appears to be \emph{anti-rotating}.

\begin{figure*}[tbh!]
 \includegraphics[width=0.33\textwidth]{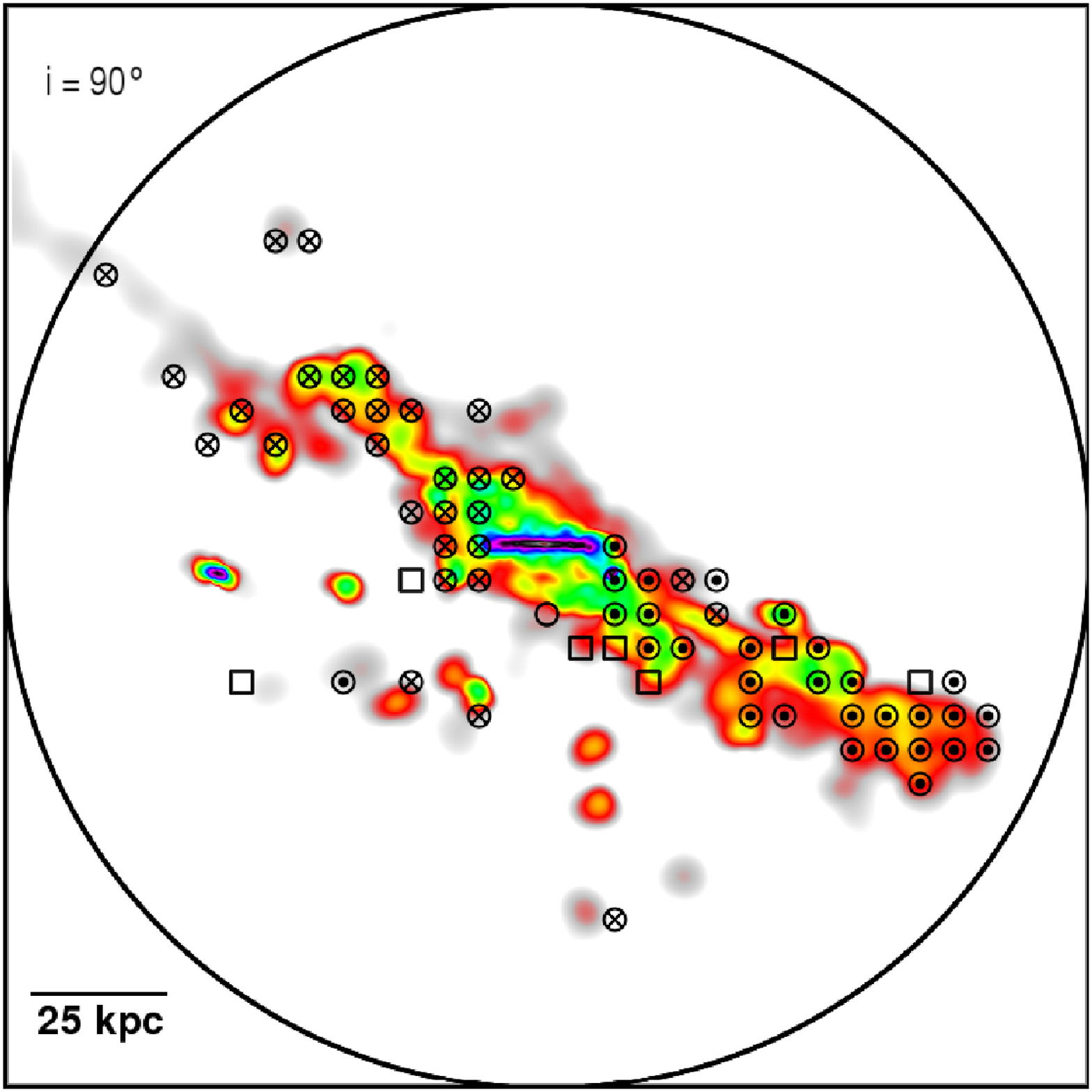}
 \includegraphics[width=0.33\textwidth]{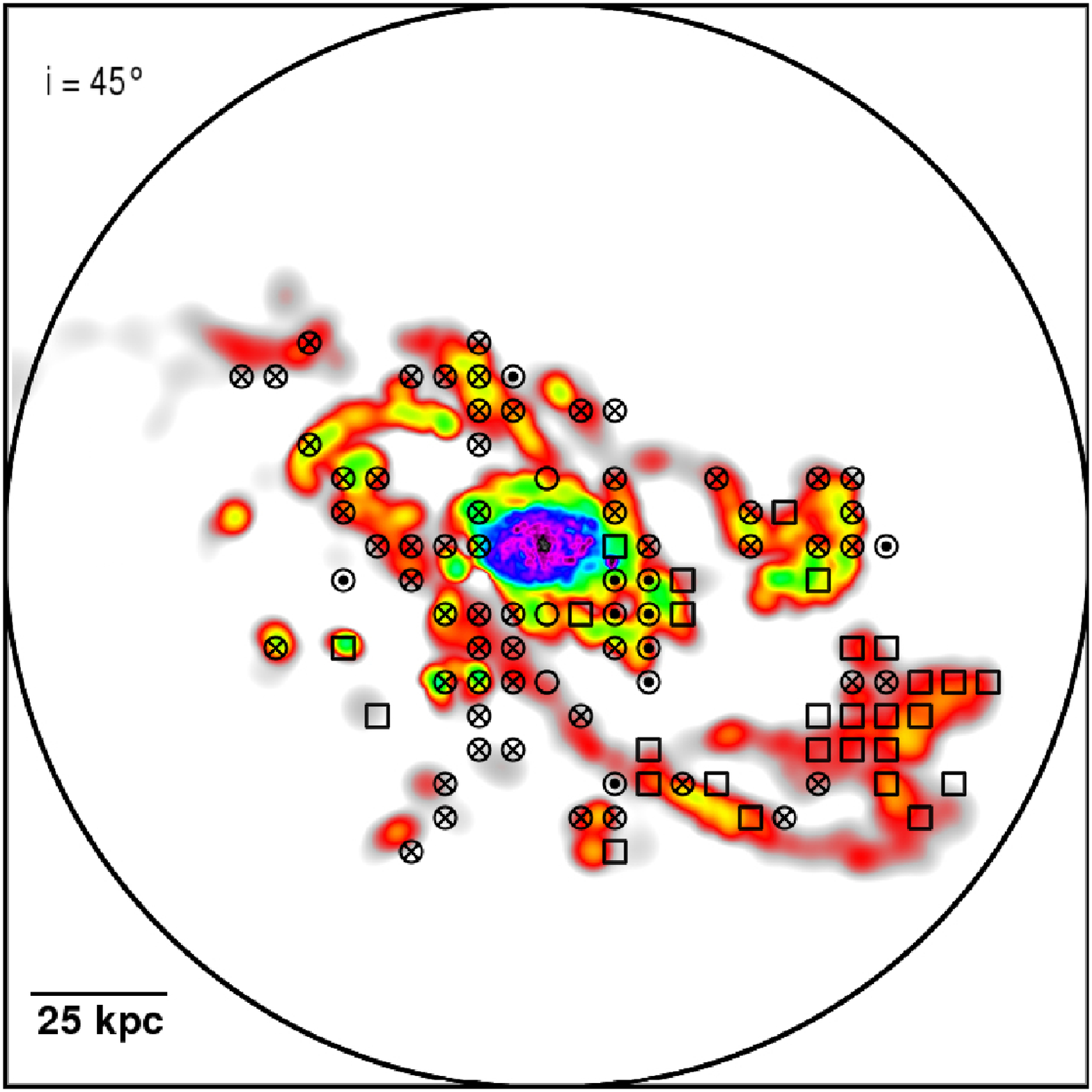}
 \includegraphics[width=0.33\textwidth]{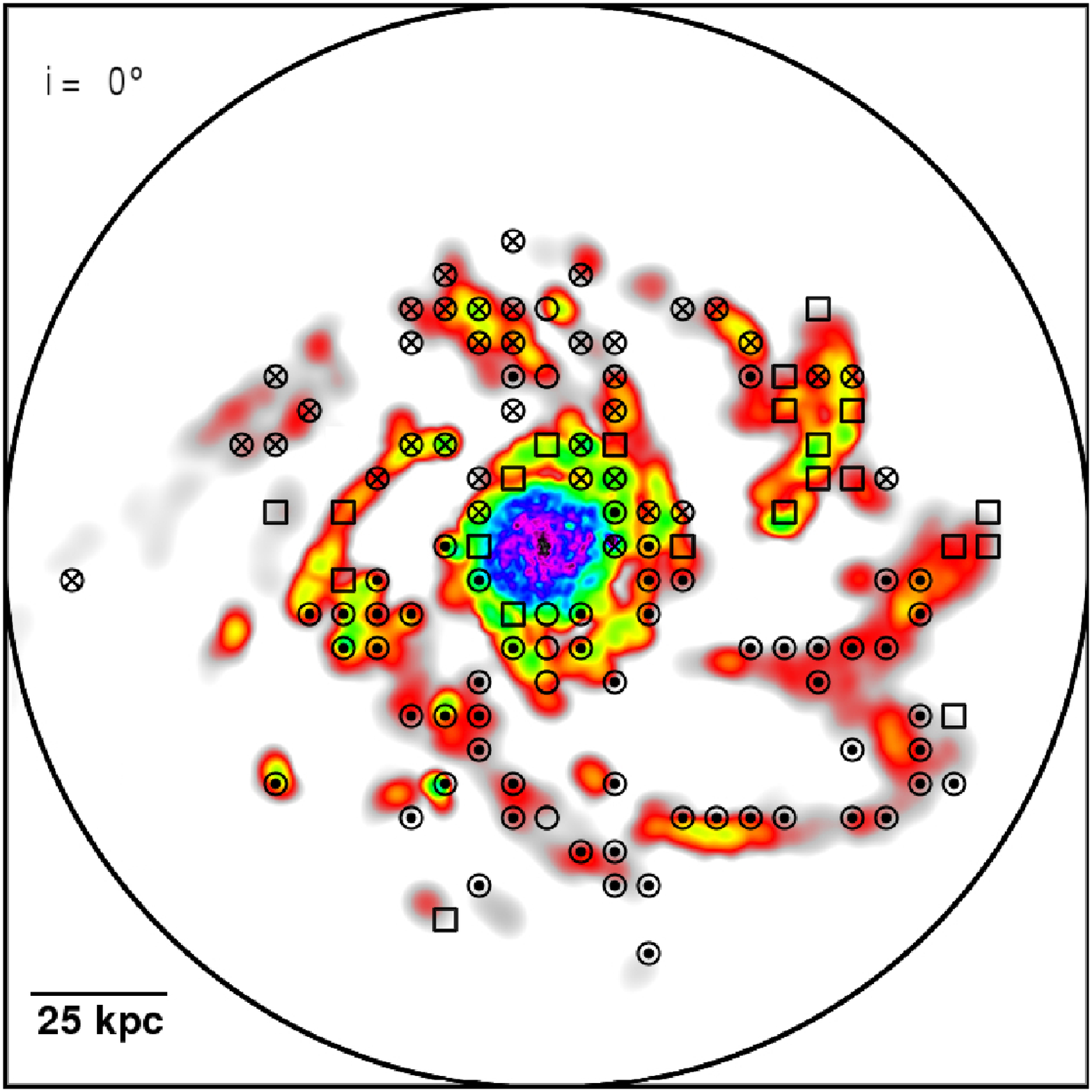}
 \includegraphics[width=0.33\textwidth]{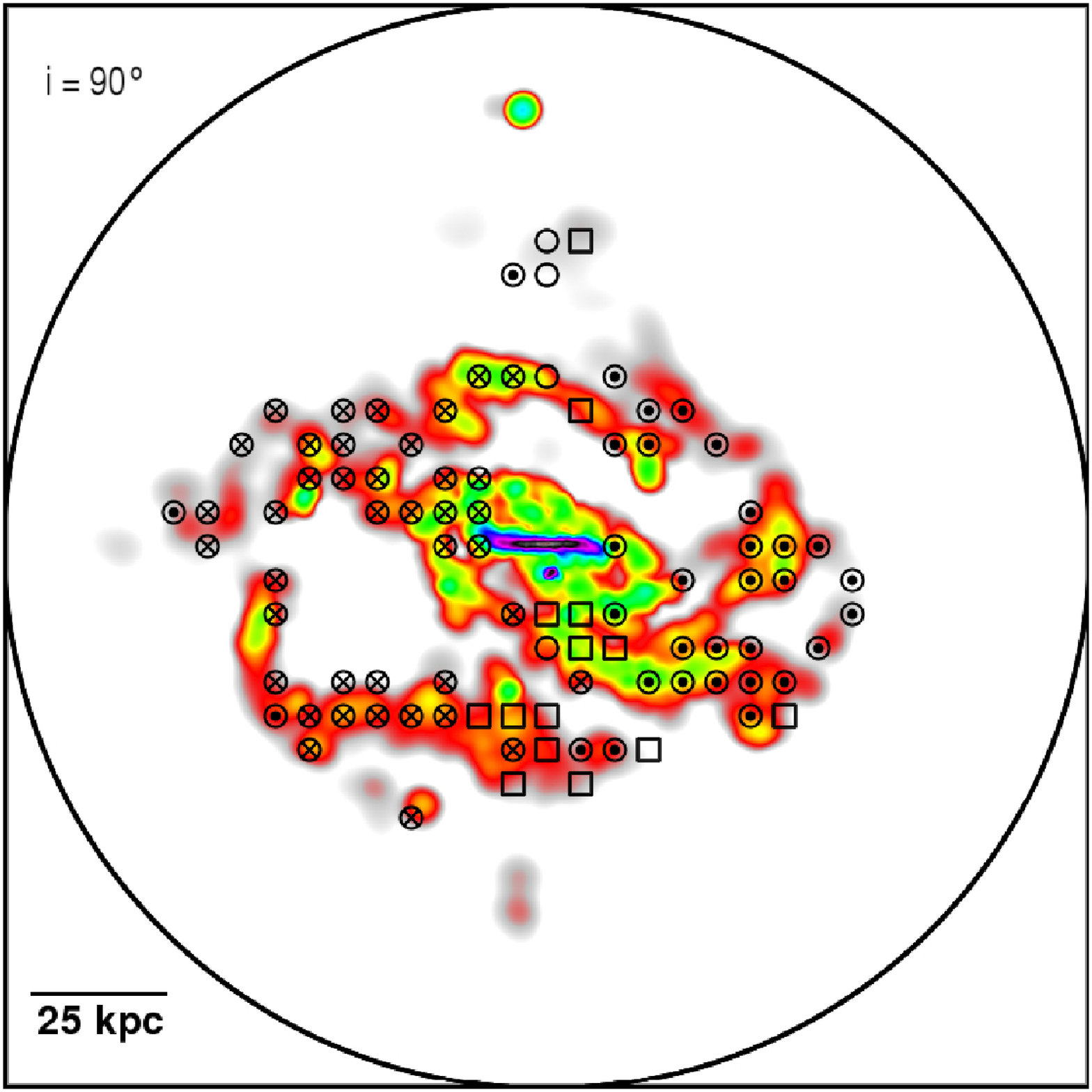}
 \includegraphics[width=0.33\textwidth]{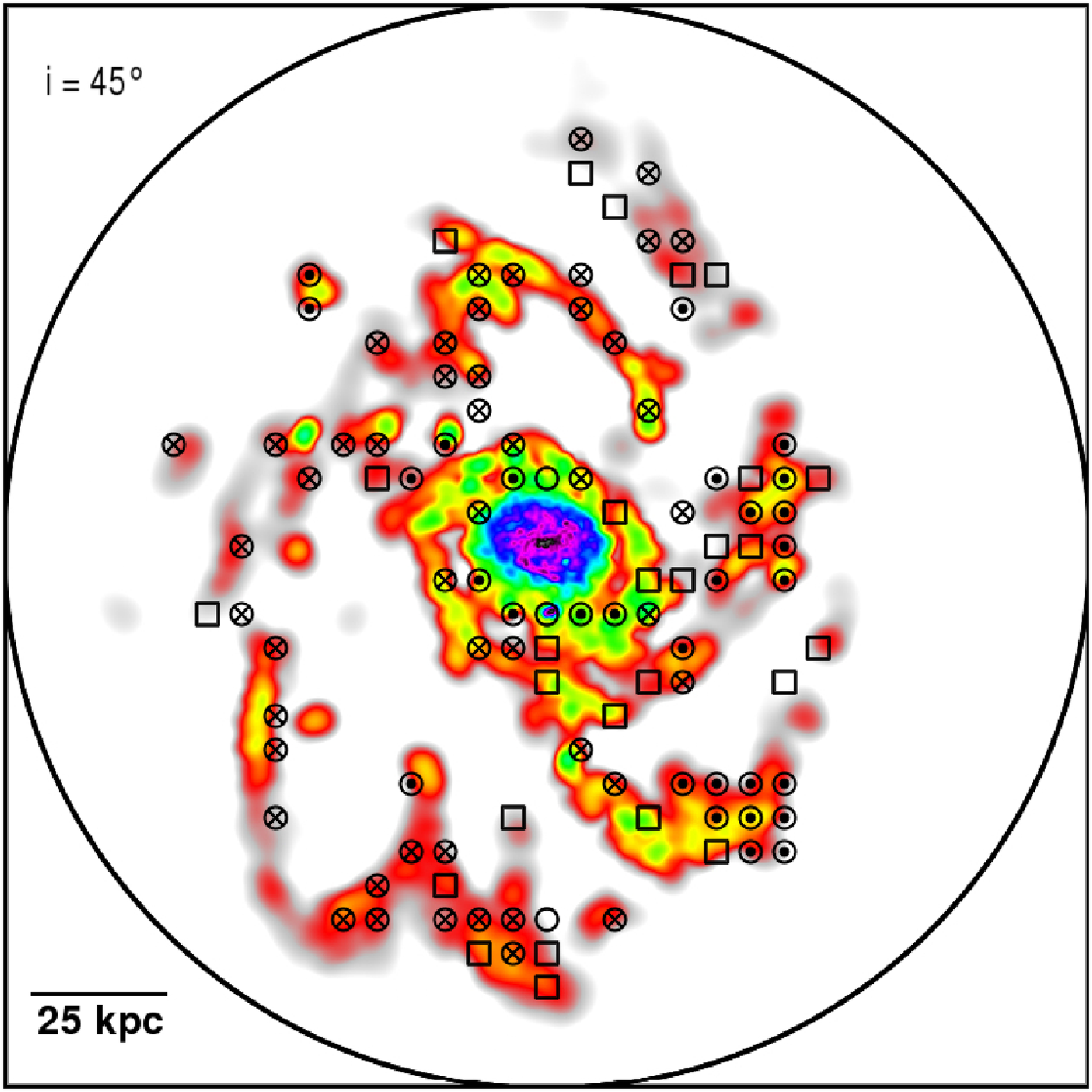}
 \includegraphics[width=0.33\textwidth]{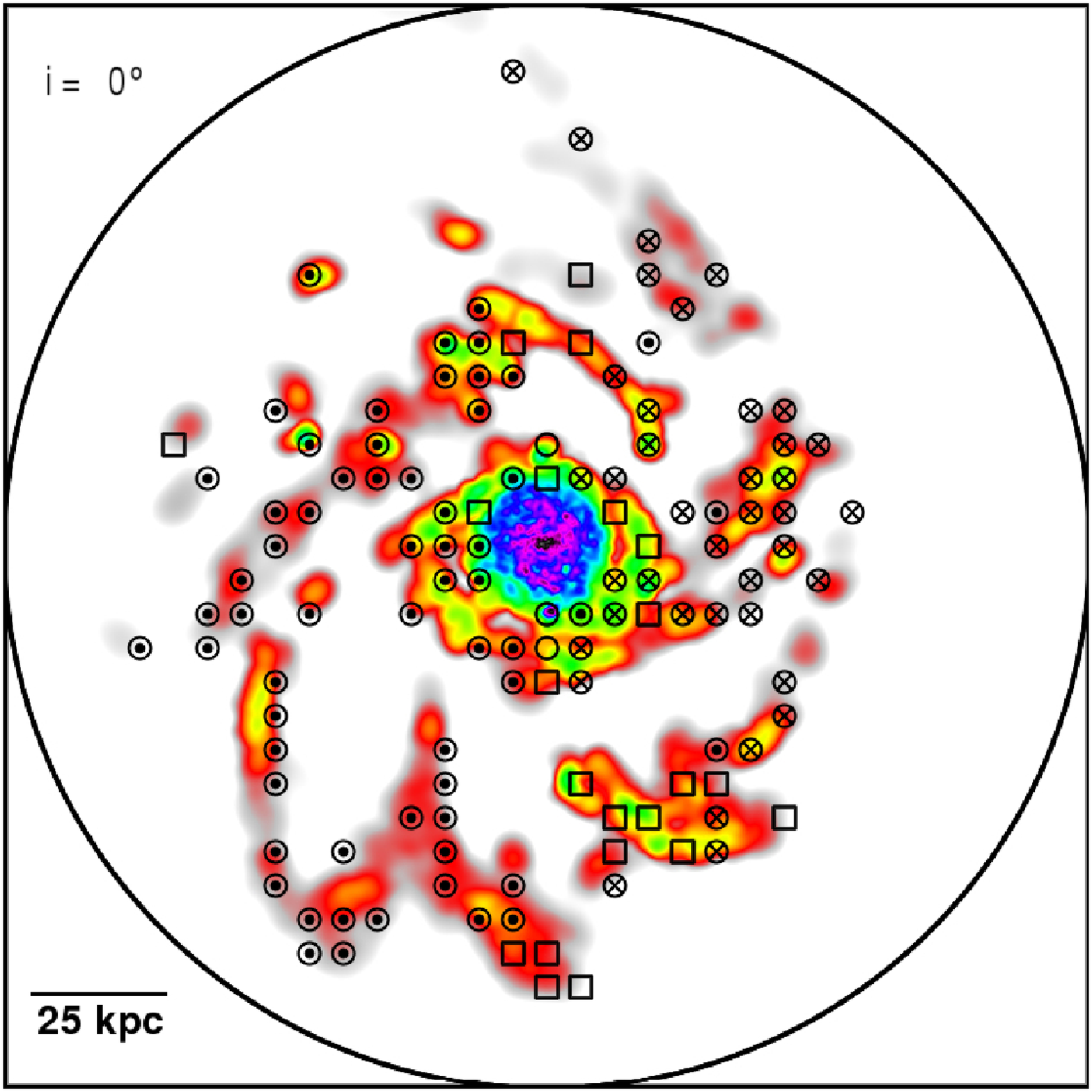}
 \caption{Identical to Figure \ref{corot_boxes}, but showing the effect of galaxy inclination on the co-rotation signature.
   The left, middle, and right panels show galaxy inclinations of $90, 45, 0$ degrees, such that the right panels are top-down
   views of the left panels.  The top and bottom panels represent
   different angles around the galaxy (at fixed inclination, $90$ degrees apart).
   Note from the top left panel that most of the cool co-rotating
   halo gas is in an extended warped disk structure.  Due to projection effects,
   gas that is co-rotating in this extended disk appears to be anti-rotating when the
   galaxy is near face-on.
   }
\label{corot_6panel}
\end{figure*}

The right panel of this figure investigates this issue further, showing the distribution of
velocity offsets for different galaxy inclinations (averaged over the four viewing angles).  When the galaxy
is viewed near edge-on (solid black line), the cool halo gas shows a wide distribution of velocities, with velocity offsets
up to $\sim300$ $\kms$ from the galaxy's systemic velocity, and an average velocity offset of
$\sim100$ $\kms$ (the vertical line).
While most sight lines still show velocity offsets at higher inclinations (i.e., most sight lines are still
orbiting) the \emph{magnitude} of these offsets are smaller, typically within $\sim100$ $\kms$ from systemic,
with an average offset of $\sim50$ $\kms$.
These smaller magnitudes allow for projection effects to grow increasingly important
for lines of sight to near face on galaxies.

In order to fully understand this trend with inclination,
Figure \ref{corot_6panel} presents six different orientations
of this galaxy (the same galaxy as Figures \ref{corot_pretty}, \ref{corot_boxes}, and \ref{corotvsangle}: Halo $2$ at $z=1.4$).  The
contours and symbols in Figure \ref{corot_6panel} are identical to Figure \ref{corot_boxes}, with the left, middle, and right panels
showing the galaxy at inclinations of $90,45,0$ degrees, respectively\footnote{The face-on panels
at the right show a ``top-down'' view of the edge-on panels at the left.}.  The top and bottom panels show two different
viewing angles around the galaxy ($90$ degrees apart).

Apparent in the upper left panel of this figure is that most of the cool halo gas around this
galaxy is confined to a single plane: a warped disk of cool accreted gas that extends to a radius of $\sim50 - 75$ co-moving
kpc \citep[not an unusual occurrence hydrodynamic simulations; e.g.][]{Agertz09,Roskar10}.
These structures, which we choose to call \emph{cold flow disks},
contribute a significant portion of the total cool halo gas in our galaxies and tend
to align roughly with the large-scale inflow filaments (see Figure \ref{zoommontage}).
Cold flow disks tend to be a common and prolonged phenomena in our simulations, and will arise naturally
in systems with the high specific angular momentum content we measure.

The top left panel demonstrates that the cold flow disk is,
in fact, co-rotating with the galaxy, since the vast majority of
absorption sight lines show co-rotation (circle-X on the left, circle-dot
on the right); i.e., the warped disk is rotating in roughly the same manner
as the galaxy.  Now consider the projection effects of viewing this same galaxy from the top-down.
Because the left side of the gaseous disk is now projected out of the image, and the right side of the gaseous disk is
projected into the image, the top half of the image is blueshifted, and the bottom half is redshifted.  Even though the gaseous
disk is co-rotating with the galaxy, the warp of the extended disk, in combination with
projection effects result in a perceived co-rotation fraction of $\sim50\%$ when the galaxy is near face-on.

The bottom panels show an even more severe example of projection effects.  In the left image, the warped disk is now seen
in projection along the image, into the image along the top half and out of the
image along the bottom half.  Again, the co-rotation signature is evident, with the left side of the warped disk
being redshifted and the right side blueshifted, the same as the galaxy.  Once this image is viewed face-on
(bottom right panel), the projected warp of the disk has completely flipped; the top half is now
projecting out of the image, and bottom half is projecting into the image.  As a result,
the line of sight velocity of the gas changes sign as well, and the entire warped disk appears to
anti-rotate with the galaxy.

As evidenced by these examples, projected co-rotation of cool halo gas
via absorption is not a reliable indicator of true co-rotation unless the galaxy is viewed near edge-on.
Even if $100\%$ of cool halo gas is in rough co-rotation with the galaxy's disk (e.g.~in a cold flow disk),
absorption lines around moderately face-on galaxies may give false impressions of anti-rotation.

\begin{figure*}[tbh!]
 \hspace{-2.5 em}
 \includegraphics[width=0.37\textwidth]{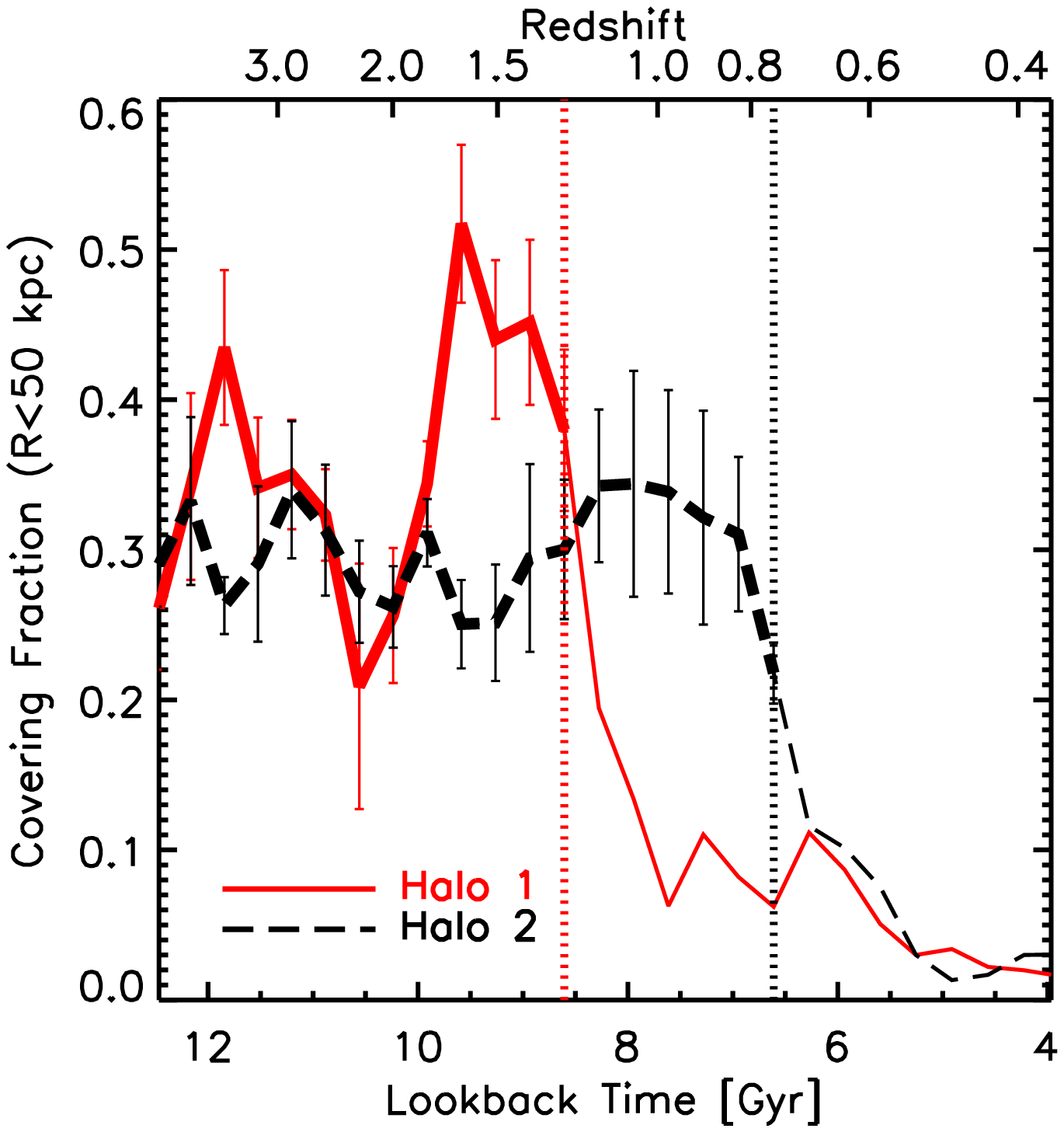}
 \hspace{-2.5 em}
 \includegraphics[width=0.37\textwidth]{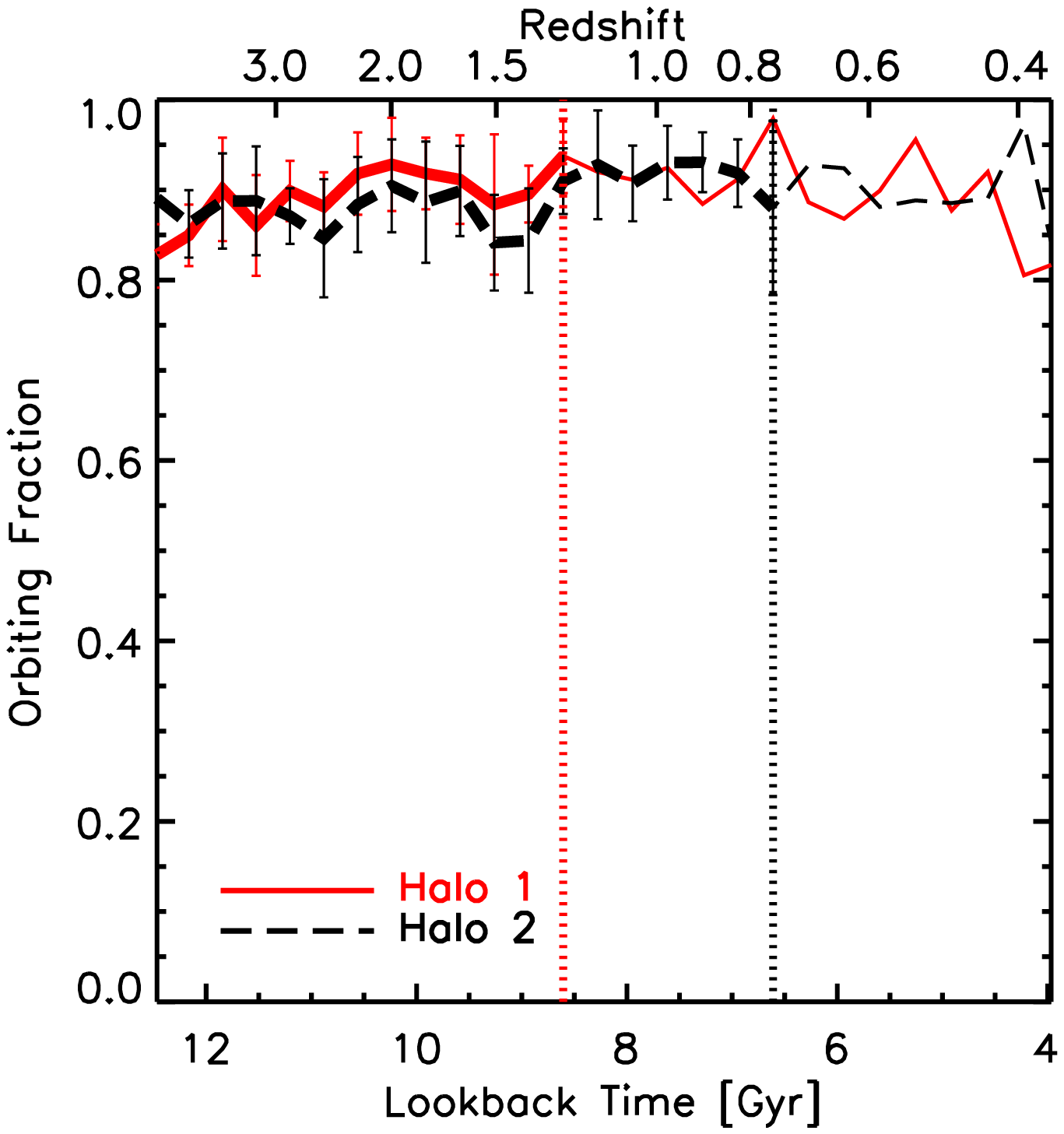}
 \hspace{-2.5 em}
 \includegraphics[width=0.37\textwidth]{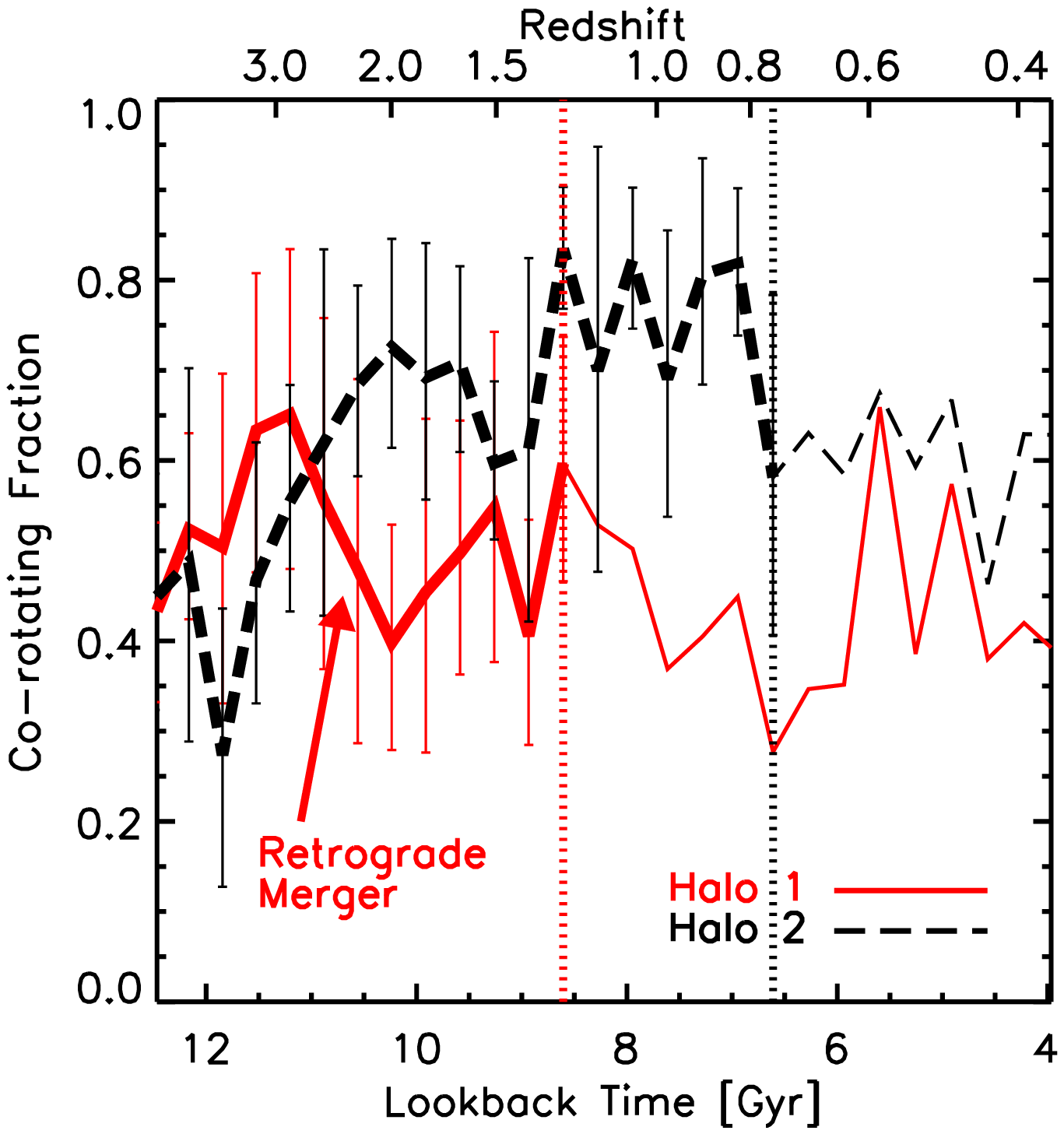}
 \caption{Covering fraction, orbiting fraction and co-rotation fraction versus
   time for both simulated galaxies, when viewed near edge-on
   (inclination $>70$ degrees).  The solid and dashed curves are for Halo $1$ and Halo $2$, respectively.
   The error bars show the $1\sigma$ scatter about the mean when averaging over $12$ different
   viewing angles (see text).
   }
\label{corotvstime}
\end{figure*}

\section{Evolution of the Orbiting and Co-rotating Signatures}
\label{vstime}
In \S\ref{kinematics}, we focused on a single example: Halo $2$ at $z=1.4$.
For this galaxy at this particular point in time, we found an overwhelming signature of orbiting gas (cool gas completely offset from
the systemic velocity if viewed in absorption) and co-rotating gas (cool gas rotating in the same direction as the galactic disk when
viewed in absorption), using mock absorption profiles along lines of sight through the galaxy halo. Our goal has been to aid direct
comparison to observations that utilize metal line absorption along lines of sight to backgrounds quasars.
We now repeat our kinematic analysis for both simulations, and at a variety of redshifts, in order to determine if our previous example
was an unusual case, or if it points to a predictable signature in the behavior of cool accreted halo gas in galaxies.
Because we found that projection effects result in unreliable rotation signatures unless the galaxy is viewed near edge-on
(in regards to co-rotation versus anti-rotation; see \S\ref{vsinclination}), we now limit our analysis to inclinations of $i\geq70$ degrees.

Figure \ref{corotvstime} shows the covering fraction (left panel),
orbiting fraction (middle panel) and co-rotating fraction (right panel) of
Halo $1$ (solid red lines) and Halo $2$ (dashed black lines) for
sight lines within $R<100$ co-moving kpc, and with $\NHI>10^{16}$ cm$^{-2}$.
Error bars show the $\sigma$ scatter due to different galaxy orientations, as discussed in \S \ref{vsinclination}.
In the right two panels, the curves change from thick to thin (and we discontinue error bars)
when the galaxy transitions from cold mode accretion to hot mode accretion
(see \S\ref{observations2} for more discussion of this transition).  After this transition, the
covering fraction of absorption gas drops precipitously (as seen in the left panel), resulting in small number statistics and large error bars.
We also note that the total mass in cool halo gas drops significantly after this transition as well (Figure \ref{masses})
and our galaxies evolve into more bulge dominated systems
(due to major mergers and secular processes, combined with the lack of a
reservoir of cool high angular momentum halo gas)
making the very definition of co-rotation with a presumed galactic disk nonsensical.
Though not shown in this figure (for clarity), we find that the results in the two right panels are nearly identical
for sight lines within $R<50$ co-moving kpc (as opposed to $R<100$), indicating a very weak dependence on radius, if any.

Although our statistics are limited to a single pair of simulated galaxies, the middle panel of Figure \ref{corotvstime}
suggests that very high orbiting fractions ($\sim80-95\%$) may be a natural consequence of cosmological gas accretion in LCDM,
even for galaxies with quite different merger histories, such as our two galaxies.
The co-rotation fraction, though, shows distinct differences between our two galaxies.
Halo $2$ experiences a quiescent merger history at $z>0.8$, and shows a very strong signature
of co-rotation once it builds a stable disk galaxy at its center ($\sim60-80\%$ of all sight lines for $z\lesssim2.5$).
Halo $1$ has a similar accretion history (and co-rotation fraction evolution) until $z\sim2.7$, at which point it
experiences a quick succession of two large mergers at $z\sim2.3$.  While mergers at high redshift
are commonplace, the infall trajectory of the merger at $z\sim2.3$ is strongly misaligned with the overall rotation of the
galaxy (and halo) at that time.  As a result, this merger completely re-orients the angular momentum axis of the galaxy, so that
Halo $1$'s galaxy at $z=1.5$ is rotating in the opposite direction as its progenitor at $z=2.5$.  Without knowing how
often major mergers result in such a drastic redistribution of angular momentum, it is impossible to speculate on
how representative Halo $1$ is, in terms of the kinematics of its cool halo gas between $z\sim1.5-2.5$.
Even in this severe case, with the angular momentum axis of the galaxy changing drastically due to a large retrograde
merger, if we disregard a brief transition period near $z=2$, Halo $1$ still shows an average co-rotation fraction
of $\sim50-60\%$ of all sight lines.

\section{Observational Implications}
\label{observations}
\subsection{Accretion Signature Versus Outflows}
\label{outflowmodels}
Summarizing our results from \S\ref{vstime}, we find that $\sim80-95\%$ of absorption sight lines in
our simulated galaxies orbit about the central galaxy, with significant velocity offsets from systemic.
The co-rotation fraction of all sight lines varies in our
two galaxies, based on their merger histories, from $\sim50-60\%$ for Halo $1$ to
$\sim60-80\%$ for Halo $2$ ($z<2.5$).
Although our lack of radiation transfer may result in quantitative changes to these results,
the qualitative kinematic behavior of our galaxies cool gaseous halos appears independent of column density.
As such, the orbiting and co-rotating signatures we have presented here should be observable via
halo absorption systems.   We note, however, that since
the co-rotation signature require galaxy kinematic and inclination information, this signature is less
useful for high redshift observations, where such information is difficult (if not impossible) to
acquire.

Given the large amount of angular momentum associated with the accreted gas in our simulations,
it is hard to imagine a scenario where radial/bi-conical outflows from galactic disks would show
such strong orbiting signatures.    In principle, gas rotating in the disk could be blown directly
upward.  In the absence of angular-momentum robbing interactions with the ambient hot-halo gas, this
gas would preserve a sense of rotation in near-polar projections from the disk.  However, the
near-polar projection is precisely where infalling gas is least likely to show velocity offsets.
One would expect these outflows to be detected along precisely those projections where infalling gas
is least likely to show velocity offsets---where the co-rotating signature
is most affected by projection effects for a near edge-on disk--—near perpendicular to the plane of the galaxy.
Of course, perfectly polar sight-lines in such a scenario would also show two-sided velocity peaks (or a broadened
peak about zero).  This is again quite different from the signature expected from accreted gas.

Moving away from the poles, the expected signal for rotating accreted gas is that a large fraction of
the sight lines within $\sim45$ degrees of the disk plane should show velocity offsets of $\sim80$ $\kms$ or more
at projected distances of $100$ co-moving kpc ($40$ physical kpc at z=1.5) from the galaxy (see Figure \ref{corotvstime}).
Consider gas that initially samples the angular momentum in a disk rotating at $\sim200$ $\kms$ and with a
half-mass radius of $\sim5$ kpc.  If that gas were blown out to projected radii of $40$ kpc without losing
any angular momentum to hot-gas halo interactions, we would expect typical velocities of only
$\sim25$ $\kms$, which is significantly smaller than what can be expected for the accreted gas.  Furthermore,
this estimate may be considered maximal in the sense that outflows are expected to preferentially expel
low angular momentum material from galaxies, rather than ejecting disk gas that uniformly samples velocities
within the disk \citep{MallerDekel02, Brook10}.

One possible concern is that the inclusion of outflowing gas may radically disrupt the behavior of the
accreted gas, such that it no longer shows a clear signature of orbiting/co-rotating about the galaxy.
While it is beyond the scope of this paper to investigate the effects of outflows on cold mode gas
accretion, a recent study by \cite{Brook10} found that while outflowing gas in their simulations (which
tended to be bi-conical, near perpendicular to the disk) had a noticeable impact on the overall dynamics
of their galaxies, the high angular momentum extended gas disks were least affected by the outflowing gas,
since these extended disks are (roughly) in the plane of the disk and outflowing material is blown
perpendicular to the disk.

\subsection{Galaxy-Absorber Kinematics}
\label{observations1}
In order to compare our findings to observations, we combine results from several
metal line absorption studies that compare the kinematics of the galaxy-absorber pairs
\citep{Steidel02, Ellison03, Chen05, Kacprzak10}.
The redshift range for the total combined sample is $0.1<z<1.0$, with typical projected radii
$R\sim20 - 100$ physical kpc between the galaxy and the absorber, and typical galaxy
luminosities of $\sim0.6 L^{\star}$.  While our simulated galaxies grow too massive
at $z\lesssim1$ to make a direct comparison in the same redshift interval
(their cold mode accretion ends, and they quickly lose their cool gaseous halos)
our galaxies are of comparable luminosity $\sim0.4L^{\star}$ at $z>1$.
While we urge the reader to keep this redshift discrepancy in mind, we believe a
useful comparison can still be made, since the covering fraction of infalling cool gas
appears to be more strongly correlated with halo mass than redshift
\citep[see Figure \ref{corotvstime}; also see][]{Stewart11a}.

Limiting these observations to galaxies with disk-like morphology
(and not counting repeat analysis of the same object), these four studies contain ($5,1,3,10$) unique pairs of objects, respectively.
Adopting our same definitions of orbiting and co-rotating sight lines
as defined in \S\ref{absorptionlines}, we find that ($4,1,2,7$) of these absorbers are orbiting,
with an observed orbiting fraction of $74\%$ ($14/19$).

Of the $14$ orbiting systems from this combined sample, ($1,0,0,1$) systems were at a near
right angle to the galactic disk.  These ambiguous systems are removed from further
comparison.  Classifying the remaining $12$ systems, we find
that ($3,1,2,3$) showed co-rotation, with only ($0,0,0,3$) showing anti-rotation.  This suggests that roughly
$75\%$ ($9/12$) of orbiting systems are co-rotating, for a total co-rotation fraction of $56\%$.
A summary comparison of these observed values to our simulations is shown in Table \ref{corottable}, including Poisson $\sqrt{N}$
errors on the total number of galaxy-absorber pairs used to compute each statistic.
In general, we find that these observations are in very good agreement with
our predicted values, suggesting that these absorption studies may constitute the first indirect observations of
cosmological gas accretion onto galaxies.
Indeed, a comparison to cosmological simulations was part of the analysis of \cite{Kacprzak10}.
They concluded that the absorption systems were consistent with the properties of cosmologically infalling gas
in LCDM.

We also compare to \citet{Kacprzak11}, who performed a similar analysis for a sample of absorber-galaxy pairs around
more massive galaxies at $z\sim0.1$ for more luminous galaxies $\sim1.9L^{\star}$.
Only $54\%$ $(7/13)$ of sight lines in this sample orbit the
galaxies\footnote{\citet{Kacprzak11} quotes a velocity offset fraction of $5/13$, rather than
$7/13$, because two of their galaxies have \emph{most} of their absorption offset to one side, but not all.
We include these two cases as orbiting for the sake of this comparison, but neither definition
changes our qualitative conclusions here.}.
The lower orbiting fraction suggests that these absorption systems show more complex kinematics than freshly
accreted gas, though the analysis of \citet{Kacprzak11} suggests that they
are not well fit by purely outflowing gas, either.
For purposes of co-rotation vs.~anti-rotation, we remove from their sample $2$
absorption lines near perpendicular to the slit across galaxy, as well as $2$ sight lines around
elliptical galaxies that are not dominated by ordered rotation.
Of this modified sample (which contains $3$ orbiting systems) only $1$ sight line shows co-rotation with the galaxy.

This co-rotation fraction is slightly lower than expected for galaxies above the
critical transition mass for shock-heating infalling gas, though
the qualitative trend agrees with our findings.
Indeed, their sample consists of galaxies with SDSS $M_r$ magnitudes in the range
(-20 to -21), appropriate for those inhabiting dark matter halos with masses that
span the transition mass $\Mvir \sim 10^{12} \Msun$ \citep[see e.g.,][]{Tollerud11}.
Unlike the less massive, higher redshift samples
(\citealp{Kacprzak10}, as well as Steidel, Ellison, Chen, and collaborators) the massive,
low-z galaxies of \citet{Kacprzak11} are \emph{not expected} to have halos dominated by freshly-accreted
(and orbiting) cold mode gas.
The paucity of cold mode accretion for massive galaxies means that the
cool gas present in these systems is more likely to be the result of outflow material
which should not display the co-rotation signatures outlined in this paper.  Alternately,
(unless it is associated visible infalling satellite galaxies) cool gas in these massive galaxies
may also be the result of cloud condensation from the hot halo,
which is roughly equally likely to be co-rotating or anti-rotating (see right panel of
Figure \ref{corotvstime}, after the shock-heating transition).  The analysis of \cite{Kacprzak11}
suggests that the systems observed are not likely to be environmental galaxy-galaxy effects, nor
from star formation driven winds, making gas accretion from the hot halo
the most likely option for this sample.

\begin{table}[tb!]
\begin{center}
\caption{ORBITING AND CO-ROTATING SIGNATURES: SIMULATIONS VS. OBSERVATIONS.}
\label{corottable}
\begin{tabular}{ | l | c | c | c |}
  \hline
                    & Observations\tablenotemark{$\dagger$}   & Halo $1$  & Halo $2$   \\
  Redshift range    & $0.1-1.0$  & $1.3-2.5$   & $0.8-2.5$ \\
  Orbiting          & $74\pm20\%$  &     $80-95\%$    & $80-95\%$       \\
  Co-rotating       & $56\pm18\%$  &     $50-60\%$    & $60-80\%$        \\ \hline
  \end{tabular}
  \vspace{-1 em}
  \end{center}
  \tablenotetext{0}{\tablenotemark{$\dagger$} Combined sample of \citet{Steidel02, Ellison03, Chen05, Kacprzak10}}
\end{table}
\vspace{1.3 em}

\subsection{Covering Fractions}
\label{observations2}
So long as galaxies experience cold mode accretion, infalling gas results in moderate covering
fractions the inner halos of these galaxies \citep{FGKeres10,Kimm10,Fumagalli11}, though the resulting
covering fractions are smaller than observed for high redshift galaxies dominated by signatures of
outflows \citep[e.g.,][]{Steidel10}.
In a previous work \citep{Stewart11a}, we examined the covering fraction
as a function of halo mass for the same two simulated galaxies studied in this paper,
finding that once each galaxy crosses a critical threshold in halo mass,
$\Mvir\sim10^{12}\Msun$, almost no cool accreted gas reaches the inner region of the halo.  The underlying cause
for this transition is because beyond a certain mass, halos are capable
of sustaining shocks in infalling gas, allowing the vast majority of cool accreted gas to shock-heat
\citep[see e.g.,][]{BirnboimDekel03,Keres05,DekelBirnboim06,Keres09, FG11, vandeVoort11}\footnote{The critical halo mass where
galaxies \emph{begin} to accrete more hot gas than cool gas is considerably lower.  Using this definition,
the transition mass is often quoted as $\sim10^{11.5}\Msun$, rather than $10^{12}\Msun$.}.
As a result, \citet{Stewart11a} found a strong trend between the covering fraction of cool halo gas and halo mass;
$\CF(<50$ co-moving kpc$) \sim30-50\%$ before this transition, and only $\sim5\%$ afterwards.  This sharp decline in
covering fraction is also shown in the left panel of Figure \ref{corotvstime}.

This transition in covering fraction should be observable in absorption studies, provided there is some way to distinguish accreted gas
from other sources of cool halo gas (e.g.~outflows from the galaxy).  The results we have presented here allow observations to
do precisely this.  For any given absorption line through a galaxy's gaseous halo, cosmologically accreted cool gas should almost always
orbit about the galaxy (orbiting fraction of $80-95\%$), and most sight lines should be co-rotating with the nearest
side of the galactic disk, as long as the galaxy is viewed near edge-on (co-rotation fraction $\sim60-70\%$).

Coupling the results of this paper with those of \citet{Stewart11a}, we predict a
clear, observable signature in the kinematics of galaxy-absorber pairs as a function of halo mass.
Below the critical halo mass, some portion of the cool
halo gas detected in absorption will be due to cool accreted
gas\footnote{We note that without knowing the full effects of outflows, it is impossible to
say how large a portion of the detectable cool gas halo will come directly from fresh accretion.}.
\emph{This accreted gas will co-rotate with the galaxy}.

Above the critical halo mass, cool accreted gas no longer contributes substantially to the covering fraction,
so even if the total covering fraction remains high (as a result of outflows) the absorbing gas will no longer co-rotate.
As a result, \emph{the covering fraction of co-rotating sight lines should drop dramatically} at or near
this critical halo mass.
While current observations lack the statistical power and halo mass range to robustly test this prediction,
we believe that future observations that continue to compare the properties of many absorption systems
with their associated galaxies will be vital in furthering our understanding
of gas accretion onto galaxies, as well as the nature of gaseous halos around galaxies.

\subsection{Connections with Extended XUV and HI Disks}
\label{warpeddisks}

The extended cold flow disk structures visualized in Figure \ref{corot_6panel} may help explain the existence of XUV
disks \citep{Thilker05,Thilker07}, large HI disks \citep[][]{GarciaRuiz02,Oosterloo07, Walter08}, and some stellar disks \citep{Barth07}
that extend to $\sim 50$ kpc (or more in some cases).
It is generally hard to understand the existence of disks with radial
extents as large as $\sim 50$ kpc in models that link disk galaxy sizes to \emph{dark matter} spins of $\lambda_{\rm dm} \sim 0.04$
\citep{FallEfst80,MoMaoWhite98,Bullock01,DuttonvandenBosch09}.  In these simple models, disk material is expected to extend
to radii $R_{\rm disk} \simeq \lambda \, R_{\rm vir} = 10$ kpc for galaxy-size halos of virial radius $R_{\rm vir} \simeq 250$ kpc.
 More extended gas disks are easier  to understand when they arise from cold-mode accretion with spin parameters as high as $\lambda \sim 0.2$
(Figure \ref{masses}).    Therefore, we expect XUV disks and extended
HI disk configurations  to be more common in halos
less massive than $\sim 10^{12} \Msun$ (or with maximum circular velocities smaller than $\sim 165$ km/s), though a detailed comparison
between the cold flow disks simulated here and observed XUV and extended HI disks is beyond the scope of this work.

\subsection{High-redshift Limitations}
While the high orbiting fractions predicted here for infalling cold mode gas only requires
a systemic velocity for the galaxy, the co-rotation signature has a notable limitation in that
it requires detailed galaxy information (rotation curve, morphology, inclination)
in order to compare galactic rotation with the associated absorption system.  While this is
possible for low redshift galaxies (e.g., the combined sample discussed previously), it is extremely difficult
at higher redshifts, making the co-rotation signature increasingly difficult to observe at high-z.

In addition, because we have focused on cool gas in galaxy halos, rather than filamentary gas that
is not associated with a particular galaxy/halo, the radial extent of this orbiting gas scales
roughly with the galaxy virial radius (this is why we have focused on co-moving coordinates in
our analysis).  As a result, the radial extent of this infalling gas (found here to be $\sim100$
co-moving kpc) corresponds to relatively small physical separations at high-z ($\sim50$ and $\sim33$
kpc at $z=1$ and $z=2$).  Once again, this is an added difficulty in observing this signature
at high redshift, as it is difficult to find bright background objects within such small projected
separations to foreground galaxies.

\section{Conclusion}
\label{conclusion}

We have used two high-resolution cosmological SPH simulations (with star formation and feedback prescriptions that have
proven effective in reproducing realistic galaxies at $z=0$) as a tool for studying the detailed properties of cool gas in
galaxy halos.  By creating absorption sight lines probing the column density of cool gas in our simulated galaxies,
we make predictions that should be comparable to observations, either via metal line absorption or direct
detection of HI, though we note that without radiative transfer, the precise values presented here are less robust than
our qualitative findings.  Our primary results are summarized as follows:

\begin{enumerate}

\item Cosmologically accreted cold-mode gas contains significant angular momentum
    compared to the dark matter, and orbits about the galactic halo before falling
    in to build the central disk.  The dimensionless spin parameter of cool halo gas
    is $\sim 3-5$ times higher than that of the dark matter, with $\lambda_{\rm cool gas} \sim 0.1-0.2$.
    Cold flows clearly supply not just baryonic mass to galaxies, but angular momentum as well.

\item This orbiting cool halo gas should be observable via background object absorption
    line studies as lines that are offset from the galaxy's systemic velocity by
    $\sim 50 - 100\,  \kms$.  Specifically,  $\sim80-95\%$ of sight lines that show absorption
    ($\NHI > 10^{16}$ cm$^{-2}$) are offset     from the central galaxy's systemic velocity
    in a single direction.     This high fraction of lines that are ``orbiting'' seems an
    unavoidable consequence of cosmological gas accretion onto disk galaxies.
    It is universal for both simulations, at all epochs, and is independent of the orientation of the galaxy.
    The orbiting gas consists of accreted cool gas from the cosmic web and subsequently falls in to help build the disk.

\item Much of the orbiting cool halo gas commonly takes the form of a fairly thick, warped, extended disk in both our galaxies.
    As a result, the fraction of all sight lines that co-rotate with the galactic disk is also relatively
    high ($\sim55\%$ for Halo $1$ and $\sim70\%$ for Halo $2$).  This is a unique, observable signature of cosmological
    gas accretion, since there is no reason to expect outflowing gas to co-rotate in this fashion.  We warn that due to line of sight projection effects of the    warped disk, this co-rotation signature can only be reliably measured if the galaxy is near edge-on ($i\gtrsim70$ degrees).  Less inclined systems may produce false anti-rotation signatures.

\item We compare our orbiting and co-rotating signatures to a combined observational sample of galaxy-absorber pairs
    for galaxies of similar luminosities
    \citep{Steidel02,Ellison03,Chen05,Kacprzak10}, finding rough agreement between our predictions and these observations
    (Table \ref{corottable}).  These studies may provide some of the first (indirect) observational evidence for
    cosmological accretion of cool gas onto galaxy halos.

\item The covering fraction of cosmologically accreted cool gas within a
    given radius, $\CFR$, is a strong function of $R$.  At small radii, the covering fraction is near unity,
    decreases to negligible values at $R>100$ co-moving kpc, and is well-fit
    by a power law.  The slope of this power law depends on the minimum column density of absorbing gas:
    $\CF(\NHI,<R)=(R/R_0)^{-\beta(\NHI)}$, with our galaxies well fit by $R_0\sim10$ co-moving kpc and
    $\beta(>10^{16}$ cm$^{-2})\sim0.6-0.8$ (see Equations 1-3).

\item The extended, warped, and thick cold-mode disks illustrated in Figure 8 may be associated with the extended UV and HI disks
    observed for galaxies in the nearby universe \citep{Thilker05,Thilker07,Oosterloo07, Walter08}.  If so, we
    expect extended disk configurations to be more common for galaxies in halos less massive than $\sim 10^{12} \Msun$ (or with maximum
    circular velocities smaller than $\sim 165$ km/s).  We note that HI and XUV disks as large as those sometimes seen ($\sim 50$ kpc)
    are quite difficult to understand if the available angular momentum for the gas mirrors that in the dark matter, as is often
    assumed in simple LCDM-based models of disk formation
    \citep[e.g.][where $R_{\rm disk} \simeq \lambda \, R_{\rm vir}$ and $\lambda \sim 0.04$]{FallEfst80, Bullock01, DuttonvandenBosch09}.
    If these disks are instead supported by cold accreted material (with $\lambda \sim 0.1-0.2$) their
    extent and structure is more naturally explained.

\end{enumerate}

Our expectation is that cold mode accretion will result in circum-galactic gas that orbits with fairly
high angular momentum.  Spherical outflows will be characteristically radial with low angular momentum,
in stark contrast to infalling gas.  While bi-conical outflows may give a velocity offsets under certain projections,
similar to the accretion signature discussed here, the resulting outflows will exhibit primarily polar configurations;
this is precisely where infalling gas is less likely to show velocity offsets and the orbiting/co-rotating signature is
least reliable.
Given a sufficiently large sample of galaxy absorber pairs, it should be possible to determine the fraction of
orbiting and non-orbiting halo gas as a function of host halo mass, and to test the expectation that less
massive halos (below the critical shock mass) should preferentially host a higher fraction of
cold-mode/orbiting gas than their more massive counterparts.

\acknowledgements
We thank Leonidas Moustakas and David Weinberg for useful discussions.
We thank the anonymous referee, whose insightful comments helped improve the quality of
this paper.  The simulations used in this paper were run on the Cosmos
cluster at JPL, and the Greenplanet cluster at UC Irvine.
This research was carried out at the Jet Propulsion Laboratory, California Institute of Technology,
under a contract with the National Aeronautics and Space Administration.
JSB and KRS were partially supported by NASA grant NNX09AG01G.
KRS is supported by an appointment to the NASA Postdoctoral Program at the Jet Propulsion Laboratory,
administered by Oak Ridge Associated Universities through a contract with NASA.
TK and JD have been supported by the Swiss National Science Foundation (SNF).
Copyright 2011. All rights reserved.

\bibliography{corot}{}

\begin{thebibliography}{97}
\expandafter\ifx\csname natexlab\endcsname\relax\def\natexlab#1{#1}\fi

\bibitem[{{Abel} {et~al.}(1997){Abel}, {Anninos}, {Zhang}, \&
  {Norman}}]{Abel97}
{Abel}, T., {Anninos}, P., {Zhang}, Y., \& {Norman}, M.~L. 1997, New Astronomy,
  2, 181, arXiv:astro-ph/9608040

\bibitem[{{Agertz} {et~al.}(2009){Agertz}, {Teyssier}, \& {Moore}}]{Agertz09}
{Agertz}, O., {Teyssier}, R., \& {Moore}, B. 2009, \mnras, 397, L64, 0901.2536

\bibitem[{{Bailin} {et~al.}(2005){Bailin}, {Kawata}, {Gibson}, {Steinmetz},
  {Navarro}, {Brook}, {Gill}, {Ibata}, {Knebe}, {Lewis}, \&
  {Okamoto}}]{Bailin05}
{Bailin}, J. {et~al.} 2005, \apjl, 627, L17, arXiv:astro-ph/0505523

\bibitem[{{Barnes} \& {Efstathiou}(1987)}]{BarnesEf87}
{Barnes}, J., \& {Efstathiou}, G. 1987, \apj, 319, 575

\bibitem[{{Barth}(2007)}]{Barth07}
{Barth}, A.~J. 2007, \aj, 133, 1085, arXiv:astro-ph/0701018

\bibitem[{{Behroozi} {et~al.}(2010){Behroozi}, {Conroy}, \&
  {Wechsler}}]{Behroozi10}
{Behroozi}, P.~S., {Conroy}, C., \& {Wechsler}, R.~H. 2010, \apj, 717, 379,
  1001.0015

\bibitem[{{Berta} {et~al.}(2008){Berta}, {Jimenez}, {Heavens}, \&
  {Panter}}]{Berta08}
{Berta}, Z.~K., {Jimenez}, R., {Heavens}, A.~F., \& {Panter}, B. 2008, \mnras,
  391, 197, 0802.1934

\bibitem[{{Bertschinger}(2001)}]{Bertschinger01}
{Bertschinger}, E. 2001, \apjs, 137, 1, arXiv:astro-ph/0103301

\bibitem[{{Bett} {et~al.}(2007){Bett}, {Eke}, {Frenk}, {Jenkins}, {Helly}, \&
  {Navarro}}]{Bett07}
{Bett}, P., {Eke}, V., {Frenk}, C.~S., {Jenkins}, A., {Helly}, J., \&
  {Navarro}, J. 2007, \mnras, 376, 215, arXiv:astro-ph/0608607

\bibitem[{{Bett} {et~al.}(2010){Bett}, {Eke}, {Frenk}, {Jenkins}, \&
  {Okamoto}}]{Bett10}
{Bett}, P., {Eke}, V., {Frenk}, C.~S., {Jenkins}, A., \& {Okamoto}, T. 2010,
  \mnras, 404, 1137, 0906.2785

\bibitem[{{Binney}(1977)}]{Binney77}
{Binney}, J. 1977, \apj, 215, 483

\bibitem[{{Birnboim} \& {Dekel}(2003)}]{BirnboimDekel03}
{Birnboim}, Y., \& {Dekel}, A. 2003, \mnras, 345, 349, arXiv:astro-ph/0302161

\bibitem[{{Black}(1981)}]{Black81}
{Black}, J.~H. 1981, \mnras, 197, 553

\bibitem[{{Blumenthal} {et~al.}(1986){Blumenthal}, {Faber}, {Flores}, \&
  {Primack}}]{Blumenthal86}
{Blumenthal}, G.~R., {Faber}, S.~M., {Flores}, R., \& {Primack}, J.~R. 1986,
  \apj, 301, 27

\bibitem[{{Bond} {et~al.}(2001){Bond}, {Churchill}, {Charlton}, \&
  {Vogt}}]{Bond01}
{Bond}, N.~A., {Churchill}, C.~W., {Charlton}, J.~C., \& {Vogt}, S.~S. 2001,
  \apj, 562, 641, arXiv:astro-ph/0108062

\bibitem[{{Bouch{\'e}} {et~al.}(2006){Bouch{\'e}}, {Murphy}, {P{\'e}roux},
  {Csabai}, \& {Wild}}]{Bouche06}
{Bouch{\'e}}, N., {Murphy}, M.~T., {P{\'e}roux}, C., {Csabai}, I., \& {Wild},
  V. 2006, \mnras, 371, 495, arXiv:astro-ph/0606328

\bibitem[{{Boylan-Kolchin} {et~al.}(2008){Boylan-Kolchin}, {Ma}, \&
  {Quataert}}]{BoylanKolchin08}
{Boylan-Kolchin}, M., {Ma}, C.-P., \& {Quataert}, E. 2008, \mnras, 383, 93,
  0707.2960

\bibitem[{{Brook} {et~al.}(2011){Brook}, {Governato}, {Ro{\v s}kar}, {Stinson},
  {Brooks}, {Wadsley}, {Quinn}, {Gibson}, {Snaith}, {Pilkington}, {House}, \&
  {Pontzen}}]{Brook10}
{Brook}, C.~B. {et~al.} 2011, \mnras, 595, 1010.1004

\bibitem[{{Brooks} {et~al.}(2007){Brooks}, {Governato}, {Booth}, {Willman},
  {Gardner}, {Wadsley}, {Stinson}, \& {Quinn}}]{Brooks07}
{Brooks}, A.~M., {Governato}, F., {Booth}, C.~M., {Willman}, B., {Gardner},
  J.~P., {Wadsley}, J., {Stinson}, G., \& {Quinn}, T. 2007, \apjl, 655, L17,
  arXiv:astro-ph/0609620

\bibitem[{{Brooks} {et~al.}(2009){Brooks}, {Governato}, {Quinn}, {Brook}, \&
  {Wadsley}}]{Brooks08}
{Brooks}, A.~M., {Governato}, F., {Quinn}, T., {Brook}, C.~B., \& {Wadsley}, J.
  2009, \apj, 694, 396, 0812.0007

\bibitem[{{Bryan} \& {Norman}(1998)}]{BryanNorman98}
{Bryan}, G.~L., \& {Norman}, M.~L. 1998, \apj, 495, 80, arXiv:astro-ph/9710107

\bibitem[{{Bullock} {et~al.}(2001){Bullock}, {Dekel}, {Kolatt}, {Kravtsov},
  {Klypin}, {Porciani}, \& {Primack}}]{Bullock01}
{Bullock}, J.~S., {Dekel}, A., {Kolatt}, T.~S., {Kravtsov}, A.~V., {Klypin},
  A.~A., {Porciani}, C., \& {Primack}, J.~R. 2001, \apj, 555, 240,
  arXiv:astro-ph/0011001

\bibitem[{{Chen} {et~al.}(2010){Chen}, {Helsby}, {Gauthier}, {Shectman},
  {Thompson}, \& {Tinker}}]{Chen10b}
{Chen}, H., {Helsby}, J.~E., {Gauthier}, J., {Shectman}, S.~A., {Thompson},
  I.~B., \& {Tinker}, J.~L. 2010, \apj, 714, 1521, 1004.0705

\bibitem[{{Chen} {et~al.}(2005){Chen}, {Kennicutt}, \& {Rauch}}]{Chen05}
{Chen}, H., {Kennicutt}, Jr., R.~C., \& {Rauch}, M. 2005, \apj, 620, 703,
  arXiv:astro-ph/0411006

\bibitem[{{Chen} \& {Tinker}(2008)}]{ChenTinker08}
{Chen}, H., \& {Tinker}, J.~L. 2008, \apj, 687, 745, 0801.2169

\bibitem[{{Churchill} {et~al.}(2005){Churchill}, {Steidel}, \&
  {Kacprzak}}]{Churchill05}
{Churchill}, C., {Steidel}, C., \& {Kacprzak}, G. 2005, in Astronomical Society
  of the Pacific Conference Series, Vol. 331, Extra-Planar Gas, ed. {R.~Braun},
  387--+

\bibitem[{{Churchill} {et~al.}(2000){Churchill}, {Mellon}, {Charlton},
  {Jannuzi}, {Kirhakos}, {Steidel}, \& {Schneider}}]{Churchill00}
{Churchill}, C.~W., {Mellon}, R.~R., {Charlton}, J.~C., {Jannuzi}, B.~T.,
  {Kirhakos}, S., {Steidel}, C.~C., \& {Schneider}, D.~P. 2000, \apjs, 130, 91,
  arXiv:astro-ph/0005585

\bibitem[{{Dekel} \& {Birnboim}(2006)}]{DekelBirnboim06}
{Dekel}, A., \& {Birnboim}, Y. 2006, \mnras, 368, 2, arXiv:astro-ph/0412300

\bibitem[{{Dekel} {et~al.}(2009){Dekel}, {Birnboim}, {Engel}, {Freundlich},
  {Goerdt}, {Mumcuoglu}, {Neistein}, {Pichon}, {Teyssier}, \&
  {Zinger}}]{Dekel09}
{Dekel}, A. {et~al.} 2009, \nat, 457, 451, 0808.0553

\bibitem[{{Diemand} {et~al.}(2008){Diemand}, {Kuhlen}, {Madau}, {Zemp},
  {Moore}, {Potter}, \& {Stadel}}]{VL2}
{Diemand}, J., {Kuhlen}, M., {Madau}, P., {Zemp}, M., {Moore}, B., {Potter},
  D., \& {Stadel}, J. 2008, \nat, 454, 735, 0805.1244

\bibitem[{{D'Onghia} \& {Navarro}(2007)}]{Donghia07}
{D'Onghia}, E., \& {Navarro}, J.~F. 2007, \mnras, 380, L58,
  arXiv:astro-ph/0703195

\bibitem[{{Dutton} \& {van den Bosch}(2009)}]{DuttonvandenBosch09}
{Dutton}, A.~A., \& {van den Bosch}, F.~C. 2009, \mnras, 396, 141, 0810.4963

\bibitem[{{Ellison} {et~al.}(2003){Ellison}, {Mall{\'e}n-Ornelas}, \&
  {Sawicki}}]{Ellison03}
{Ellison}, S.~L., {Mall{\'e}n-Ornelas}, G., \& {Sawicki}, M. 2003, \apj, 589,
  709, arXiv:astro-ph/0302147

\bibitem[{{Fall} \& {Efstathiou}(1980)}]{FallEfst80}
{Fall}, S.~M., \& {Efstathiou}, G. 1980, \mnras, 193, 189

\bibitem[{{Faucher-Giguere} {et~al.}(2011){Faucher-Giguere}, {Keres}, \&
  {Ma}}]{FG11}
{Faucher-Giguere}, C., {Keres}, D., \& {Ma}, C. 2011, ArXiv e-prints, 1103.0001

\bibitem[{{Faucher-Gigu{\`e}re} \& {Kere{\v s}}(2011)}]{FGKeres10}
{Faucher-Gigu{\`e}re}, C.-A., \& {Kere{\v s}}, D. 2011, \mnras, 412, L118,
  1011.1693

\bibitem[{{Fumagalli} {et~al.}(2011){Fumagalli}, {Prochaska}, {Kasen}, {Dekel},
  {Ceverino}, \& {Primack}}]{Fumagalli11}
{Fumagalli}, M., {Prochaska}, J.~X., {Kasen}, D., {Dekel}, A., {Ceverino}, D.,
  \& {Primack}, J.~R. 2011, ArXiv e-prints, 1103.2130

\bibitem[{{Garc{\'{\i}}a-Ruiz} {et~al.}(2002){Garc{\'{\i}}a-Ruiz}, {Sancisi},
  \& {Kuijken}}]{GarciaRuiz02}
{Garc{\'{\i}}a-Ruiz}, I., {Sancisi}, R., \& {Kuijken}, K. 2002, \aap, 394, 769,
  arXiv:astro-ph/0207112

\bibitem[{{Governato} {et~al.}(2010){Governato}, {Brook}, {Mayer}, {Brooks},
  {Rhee}, {Wadsley}, {Jonsson}, {Willman}, {Stinson}, {Quinn}, \&
  {Madau}}]{Governato10}
{Governato}, F. {et~al.} 2010, \nat, 463, 203

\bibitem[{{Governato} {et~al.}(2009){Governato}, {Brook}, {Brooks}, {Mayer},
  {Willman}, {Jonsson}, {Stilp}, {Pope}, {Christensen}, {Wadsley}, \&
  {Quinn}}]{Governato08}
------. 2009, \mnras, 398, 312, 0812.0379

\bibitem[{{Governato} {et~al.}(2007){Governato}, {Willman}, {Mayer}, {Brooks},
  {Stinson}, {Valenzuela}, {Wadsley}, \& {Quinn}}]{Governato07}
{Governato}, F., {Willman}, B., {Mayer}, L., {Brooks}, A., {Stinson}, G.,
  {Valenzuela}, O., {Wadsley}, J., \& {Quinn}, T. 2007, \mnras, 374, 1479,
  arXiv:astro-ph/0602351

\bibitem[{{Guo} {et~al.}(2010){Guo}, {White}, {Li}, \&
  {Boylan-Kolchin}}]{Guo10}
{Guo}, Q., {White}, S., {Li}, C., \& {Boylan-Kolchin}, M. 2010, \mnras, 404,
  1111, 0909.4305

\bibitem[{{Haardt} \& {Madau}(1996)}]{HaardtMadau96}
{Haardt}, F., \& {Madau}, P. 1996, \apj, 461, 20, arXiv:astro-ph/9509093

\bibitem[{{Hopkins} \& {Beacom}(2006)}]{HopkinsBeacom06}
{Hopkins}, A.~M., \& {Beacom}, J.~F. 2006, \apj, 651, 142,
  arXiv:astro-ph/0601463

\bibitem[{{Kacprzak} {et~al.}(2011){Kacprzak}, {Churchill}, {Barton}, \&
  {Cooke}}]{Kacprzak11}
{Kacprzak}, G.~G., {Churchill}, C.~W., {Barton}, E.~J., \& {Cooke}, J. 2011,
  \apj, 733, 105, 1102.4339

\bibitem[{{Kacprzak} {et~al.}(2010){Kacprzak}, {Churchill}, {Ceverino},
  {Steidel}, {Klypin}, \& {Murphy}}]{Kacprzak10}
{Kacprzak}, G.~G., {Churchill}, C.~W., {Ceverino}, D., {Steidel}, C.~C.,
  {Klypin}, A., \& {Murphy}, M.~T. 2010, \apj, 711, 533, 0912.2746

\bibitem[{{Kacprzak} {et~al.}(2007){Kacprzak}, {Churchill}, {Steidel},
  {Murphy}, \& {Evans}}]{Kacprzak07}
{Kacprzak}, G.~G., {Churchill}, C.~W., {Steidel}, C.~C., {Murphy}, M.~T., \&
  {Evans}, J.~L. 2007, \apj, 662, 909, arXiv:astro-ph/0703377

\bibitem[{{Katz} {et~al.}(1996){Katz}, {Weinberg}, \& {Hernquist}}]{Katz96}
{Katz}, N., {Weinberg}, D.~H., \& {Hernquist}, L. 1996, \apjs, 105, 19,
  arXiv:astro-ph/9509107

\bibitem[{{Kere{\v s}} \& {Hernquist}(2009)}]{KeresHernquist09}
{Kere{\v s}}, D., \& {Hernquist}, L. 2009, \apjl, 700, L1, 0905.2186

\bibitem[{{Kere{\v s}} {et~al.}(2009{\natexlab{a}}){Kere{\v s}}, {Katz},
  {Dav{\'e}}, {Fardal}, \& {Weinberg}}]{Keres09b}
{Kere{\v s}}, D., {Katz}, N., {Dav{\'e}}, R., {Fardal}, M., \& {Weinberg},
  D.~H. 2009{\natexlab{a}}, \mnras, 396, 2332, 0901.1880

\bibitem[{{Kere{\v s}} {et~al.}(2009{\natexlab{b}}){Kere{\v s}}, {Katz},
  {Fardal}, {Dav{\'e}}, \& {Weinberg}}]{Keres09}
{Kere{\v s}}, D., {Katz}, N., {Fardal}, M., {Dav{\'e}}, R., \& {Weinberg},
  D.~H. 2009{\natexlab{b}}, \mnras, 395, 160, 0809.1430

\bibitem[{{Kere{\v s}} {et~al.}(2005){Kere{\v s}}, {Katz}, {Weinberg}, \&
  {Dav{\'e}}}]{Keres05}
{Kere{\v s}}, D., {Katz}, N., {Weinberg}, D.~H., \& {Dav{\'e}}, R. 2005,
  \mnras, 363, 2, arXiv:astro-ph/0407095

\bibitem[{{Kimm} {et~al.}(2011){Kimm}, {Slyz}, {Devriendt}, \&
  {Pichon}}]{Kimm10}
{Kimm}, T., {Slyz}, A., {Devriendt}, J., \& {Pichon}, C. 2011, \mnras, 413,
  L51, 1012.0059

\bibitem[{{Macci{\`o}} {et~al.}(2007){Macci{\`o}}, {Dutton}, {van den Bosch},
  {Moore}, {Potter}, \& {Stadel}}]{Maccio07}
{Macci{\`o}}, A.~V., {Dutton}, A.~A., {van den Bosch}, F.~C., {Moore}, B.,
  {Potter}, D., \& {Stadel}, J. 2007, \mnras, 378, 55, arXiv:astro-ph/0608157

\bibitem[{{Maller} \& {Bullock}(2004)}]{MallerBullock04}
{Maller}, A.~H., \& {Bullock}, J.~S. 2004, \mnras, 355, 694,
  arXiv:astro-ph/0406632

\bibitem[{{Maller} \& {Dekel}(2002)}]{MallerDekel02}
{Maller}, A.~H., \& {Dekel}, A. 2002, \mnras, 335, 487, arXiv:astro-ph/0201187

\bibitem[{{Maller} {et~al.}(2002){Maller}, {Dekel}, \& {Somerville}}]{Maller02}
{Maller}, A.~H., {Dekel}, A., \& {Somerville}, R. 2002, \mnras, 329, 423,
  arXiv:astro-ph/0105168

\bibitem[{{Martin}(2005)}]{Martin05}
{Martin}, C.~L. 2005, \apj, 621, 227, arXiv:astro-ph/0410247

\bibitem[{{Martin} \& {Bouch{\'e}}(2009)}]{MartinBouche09}
{Martin}, C.~L., \& {Bouch{\'e}}, N. 2009, \apj, 703, 1394, 0908.4271

\bibitem[{{Mo} {et~al.}(1998){Mo}, {Mao}, \& {White}}]{MoMaoWhite98}
{Mo}, H.~J., {Mao}, S., \& {White}, S.~D.~M. 1998, \mnras, 295, 319,
  arXiv:astro-ph/9707093

\bibitem[{{Oosterloo} {et~al.}(2007){Oosterloo}, {Morganti}, {Sadler}, {van der
  Hulst}, \& {Serra}}]{Oosterloo07}
{Oosterloo}, T.~A., {Morganti}, R., {Sadler}, E.~M., {van der Hulst}, T., \&
  {Serra}, P. 2007, \aap, 465, 787, arXiv:astro-ph/0701716

\bibitem[{{Peebles}(1969)}]{Peebles69}
{Peebles}, P.~J.~E. 1969, \apj, 155, 393

\bibitem[{{Pontzen} {et~al.}(2008){Pontzen}, {Governato}, {Pettini}, {Booth},
  {Stinson}, {Wadsley}, {Brooks}, {Quinn}, \& {Haehnelt}}]{Pontzen08}
{Pontzen}, A. {et~al.} 2008, \mnras, 390, 1349, 0804.4474

\bibitem[{{Prochter} {et~al.}(2006){Prochter}, {Prochaska}, \&
  {Burles}}]{Prochter06}
{Prochter}, G.~E., {Prochaska}, J.~X., \& {Burles}, S.~M. 2006, \apj, 639, 766,
  arXiv:astro-ph/0411776

\bibitem[{{Rigby} {et~al.}(2002){Rigby}, {Charlton}, \& {Churchill}}]{Rigby02}
{Rigby}, J.~R., {Charlton}, J.~C., \& {Churchill}, C.~W. 2002, \apj, 565, 743,
  arXiv:astro-ph/0110191

\bibitem[{{Ro{\v s}kar} {et~al.}(2010){Ro{\v s}kar}, {Debattista}, {Brooks},
  {Quinn}, {Brook}, {Governato}, {Dalcanton}, \& {Wadsley}}]{Roskar10}
{Ro{\v s}kar}, R., {Debattista}, V.~P., {Brooks}, A.~M., {Quinn}, T.~R.,
  {Brook}, C.~B., {Governato}, F., {Dalcanton}, J.~J., \& {Wadsley}, J. 2010,
  \mnras, 408, 783, 1006.1659

\bibitem[{{Rubin} {et~al.}(2010){Rubin}, {Prochaska}, {Koo}, {Phillips}, \&
  {Weiner}}]{Rubin10}
{Rubin}, K.~H.~R., {Prochaska}, J.~X., {Koo}, D.~C., {Phillips}, A.~C., \&
  {Weiner}, B.~J. 2010, \apj, 712, 574, 0907.0231

\bibitem[{{Rubin} {et~al.}(2011){Rubin}, {Prochaska}, {M{\'e}nard}, {Murray},
  {Kasen}, {Koo}, \& {Phillips}}]{Rubin11}
{Rubin}, K.~H.~R., {Prochaska}, J.~X., {M{\'e}nard}, B., {Murray}, N., {Kasen},
  D., {Koo}, D.~C., \& {Phillips}, A.~C. 2011, \apj, 728, 55, 1008.3397

\bibitem[{{Shapley} {et~al.}(2003){Shapley}, {Steidel}, {Pettini}, \&
  {Adelberger}}]{Shapley03}
{Shapley}, A.~E., {Steidel}, C.~C., {Pettini}, M., \& {Adelberger}, K.~L. 2003,
  \apj, 588, 65, arXiv:astro-ph/0301230

\bibitem[{{Sharma} \& {Steinmetz}(2005)}]{SharmaSteinmetz05}
{Sharma}, S., \& {Steinmetz}, M. 2005, \apj, 628, 21, arXiv:astro-ph/0406533

\bibitem[{{Shen} {et~al.}(2010){Shen}, {Wadsley}, \& {Stinson}}]{Shen10}
{Shen}, S., {Wadsley}, J., \& {Stinson}, G. 2010, \mnras, 407, 1581, 0910.5956

\bibitem[{{Silk}(1977)}]{Silk77}
{Silk}, J. 1977, \apj, 211, 638

\bibitem[{{Spergel} {et~al.}(2007){Spergel}, {Bean}, {Dor{\'e}}, {Nolta},
  {Bennett}, {Dunkley}, {Hinshaw}, {Jarosik}, {Komatsu}, {Page}, {Peiris},
  {Verde}, {Halpern}, {Hill}, {Kogut}, {Limon}, {Meyer}, {Odegard}, {Tucker},
  {Weiland}, {Wollack}, \& {Wright}}]{WMAP3}
{Spergel}, D.~N. {et~al.} 2007, \apjs, 170, 377, arXiv:astro-ph/0603449

\bibitem[{{Stadel}(2001)}]{Stadel01}
{Stadel}, J.~G. 2001, PhD thesis, University of Washington

\bibitem[{{Steidel}(1995)}]{Steidel95}
{Steidel}, C.~C. 1995, in QSO Absorption Lines, ed. {G.~Meylan}, 139--+

\bibitem[{{Steidel} {et~al.}(2010){Steidel}, {Erb}, {Shapley}, {Pettini},
  {Reddy}, {Bogosavljevi{\'c}}, {Rudie}, \& {Rakic}}]{Steidel10}
{Steidel}, C.~C., {Erb}, D.~K., {Shapley}, A.~E., {Pettini}, M., {Reddy}, N.,
  {Bogosavljevi{\'c}}, M., {Rudie}, G.~C., \& {Rakic}, O. 2010, \apj, 717, 289,
  1003.0679

\bibitem[{{Steidel} {et~al.}(1996){Steidel}, {Giavalisco}, {Pettini},
  {Dickinson}, \& {Adelberger}}]{Steidel96}
{Steidel}, C.~C., {Giavalisco}, M., {Pettini}, M., {Dickinson}, M., \&
  {Adelberger}, K.~L. 1996, \apjl, 462, L17+, arXiv:astro-ph/9602024

\bibitem[{{Steidel} {et~al.}(2002){Steidel}, {Kollmeier}, {Shapley},
  {Churchill}, {Dickinson}, \& {Pettini}}]{Steidel02}
{Steidel}, C.~C., {Kollmeier}, J.~A., {Shapley}, A.~E., {Churchill}, C.~W.,
  {Dickinson}, M., \& {Pettini}, M. 2002, \apj, 570, 526,
  arXiv:astro-ph/0201353

\bibitem[{{Stewart} {et~al.}(2009{\natexlab{a}}){Stewart}, {Bullock}, {Barton},
  \& {Wechsler}}]{Stewart09a}
{Stewart}, K.~R., {Bullock}, J.~S., {Barton}, E.~J., \& {Wechsler}, R.~H.
  2009{\natexlab{a}}, \apj, 702, 1005, 0811.1218

\bibitem[{{Stewart} {et~al.}(2009{\natexlab{b}}){Stewart}, {Bullock},
  {Wechsler}, \& {Maller}}]{Stewart09b}
{Stewart}, K.~R., {Bullock}, J.~S., {Wechsler}, R.~H., \& {Maller}, A.~H.
  2009{\natexlab{b}}, \apj, 702, 307, 0901.4336

\bibitem[{{Stewart} {et~al.}(2008){Stewart}, {Bullock}, {Wechsler}, {Maller},
  \& {Zentner}}]{Stewart08}
{Stewart}, K.~R., {Bullock}, J.~S., {Wechsler}, R.~H., {Maller}, A.~H., \&
  {Zentner}, A.~R. 2008, \apj, 683, 597

\bibitem[{{Stewart} {et~al.}(2011){Stewart}, {Kaufmann}, {Bullock}, {Barton},
  {Maller}, {Diemand}, \& {Wadsley}}]{Stewart11a}
{Stewart}, K.~R., {Kaufmann}, T., {Bullock}, J.~S., {Barton}, E.~J., {Maller},
  A.~H., {Diemand}, J., \& {Wadsley}, J. 2011, \apjl, 735, L1+, 1012.2128

\bibitem[{{Stinson et. al}(2006)}]{Stinson06}
{Stinson et. al}. 2006, \mnras, 373, 1074, arXiv:astro-ph/0602350

\bibitem[{{Thilker} {et~al.}(2005){Thilker}, {Bianchi}, {Boissier}, {Gil de
  Paz}, {Madore}, {Martin}, {Meurer}, {Neff}, {Rich}, {Schiminovich},
  {Seibert}, {Wyder}, {Barlow}, {Byun}, {Donas}, {Forster}, {Friedman},
  {Heckman}, {Jelinsky}, {Lee}, {Malina}, {Milliard}, {Morrissey}, {Siegmund},
  {Small}, {Szalay}, \& {Welsh}}]{Thilker05}
{Thilker}, D.~A. {et~al.} 2005, \apjl, 619, L79, arXiv:astro-ph/0411306

\bibitem[{{Thilker} {et~al.}(2007){Thilker}, {Bianchi}, {Meurer}, {Gil de Paz},
  {Boissier}, {Madore}, {Boselli}, {Ferguson}, {Mu{\~n}oz-Mateos}, {Madsen},
  {Hameed}, {Overzier}, {Forster}, {Friedman}, {Martin}, {Morrissey}, {Neff},
  {Schiminovich}, {Seibert}, {Small}, {Wyder}, {Donas}, {Heckman}, {Lee},
  {Milliard}, {Rich}, {Szalay}, {Welsh}, \& {Yi}}]{Thilker07}
------. 2007, \apjs, 173, 538, 0712.3555

\bibitem[{{Tinker} \& {Chen}(2008)}]{TinkerChen08}
{Tinker}, J.~L., \& {Chen}, H. 2008, \apj, 679, 1218, 0709.1470

\bibitem[{{Tollerud} {et~al.}(2011){Tollerud}, {Bullock}, {Graves}, \&
  {Wolf}}]{Tollerud11}
{Tollerud}, E.~J., {Bullock}, J.~S., {Graves}, G.~J., \& {Wolf}, J. 2011, \apj,
  726, 108, 1007.5311

\bibitem[{{van de Voort} {et~al.}(2011){van de Voort}, {Schaye}, {Booth}, \&
  {Dalla Vecchia}}]{vandeVoort11}
{van de Voort}, F., {Schaye}, J., {Booth}, C.~M., \& {Dalla Vecchia}, C. 2011,
  \mnras, 805, 1102.3912

\bibitem[{{van den Bosch} {et~al.}(2002){van den Bosch}, {Abel}, {Croft},
  {Hernquist}, \& {White}}]{vandenBosch02}
{van den Bosch}, F.~C., {Abel}, T., {Croft}, R.~A.~C., {Hernquist}, L., \&
  {White}, S.~D.~M. 2002, \apj, 576, 21, arXiv:astro-ph/0201095

\bibitem[{{Verner} \& {Ferland}(1996)}]{Verner96}
{Verner}, D.~A., \& {Ferland}, G.~J. 1996, \apjs, 103, 467,
  arXiv:astro-ph/9509083

\bibitem[{{Vitvitska} {et~al.}(2002){Vitvitska}, {Klypin}, {Kravtsov},
  {Wechsler}, {Primack}, \& {Bullock}}]{Vitvitska02}
{Vitvitska}, M., {Klypin}, A.~A., {Kravtsov}, A.~V., {Wechsler}, R.~H.,
  {Primack}, J.~R., \& {Bullock}, J.~S. 2002, \apj, 581, 799,
  arXiv:astro-ph/0105349

\bibitem[{{Wadsley et. al}(2004)}]{GASOLINE}
{Wadsley et. al}. 2004, New Astronomy, 9, 137, arXiv:astro-ph/0303521

\bibitem[{{Walter} {et~al.}(2008){Walter}, {Brinks}, {de Blok}, {Bigiel},
  {Kennicutt}, {Thornley}, \& {Leroy}}]{Walter08}
{Walter}, F., {Brinks}, E., {de Blok}, W.~J.~G., {Bigiel}, F., {Kennicutt},
  R.~C., {Thornley}, M.~D., \& {Leroy}, A. 2008, \aj, 136, 2563, 0810.2125

\bibitem[{{Weiner} {et~al.}(2009){Weiner}, {Coil}, {Prochaska}, {Newman},
  {Cooper}, {Bundy}, {Conselice}, {Dutton}, {Faber}, {Koo}, {Lotz}, {Rieke}, \&
  {Rubin}}]{Weiner09}
{Weiner}, B.~J. {et~al.} 2009, \apj, 692, 187, 0804.4686

\bibitem[{{Weinmann} {et~al.}(2009){Weinmann}, {Kauffmann}, {van den Bosch},
  {Pasquali}, {McIntosh}, {Mo}, {Yang}, \& {Guo}}]{Weinmann09}
{Weinmann}, S.~M., {Kauffmann}, G., {van den Bosch}, F.~C., {Pasquali}, A.,
  {McIntosh}, D.~H., {Mo}, H., {Yang}, X., \& {Guo}, Y. 2009, \mnras, 394,
  1213, 0809.2283

\bibitem[{{White} \& {Frenk}(1991)}]{WhiteFrenk91}
{White}, S.~D.~M., \& {Frenk}, C.~S. 1991, \apj, 379, 52

\bibitem[{{White} \& {Rees}(1978)}]{WhiteRees78}
{White}, S.~D.~M., \& {Rees}, M.~J. 1978, \mnras, 183, 341

\end{thebibliography}
\bibliographystyle{hapj}

\end{document}